\documentclass[usenatbib]{mnras_PaperI}
\usepackage[fleqn]{amsmath}
\usepackage{array,multirow,graphicx}
\usepackage{txfonts}
\usepackage{lmodern}
\bibpunct{(}{)}{;}{a}{}{,}
\usepackage{floatrow}
\usepackage{array}
\usepackage{relsize}
\usepackage{bm}
\usepackage{listings}
\usepackage{url}
\usepackage[font=small,labelfont=bf]{caption}

\graphicspath{{./figs/}}

\newcommand{\zo}{$z\!=\!0$}
\newcommand{\HI}{H\,{\sc i}} 
\newcommand{\Htwo}{H$_{2}$}

\newfloatcommand{capbtabbox}{table}[][\FBwidth]

\def\app#1#2{%
\mathrel{%
\setbox0=\hbox{$#1\sim$}%
\setbox2=\hbox{%
\rlap{\hbox{$#1\propto$}}%
\lower1.1\ht0\box0%
}%
\raise0.25\ht2\box2%
}%
}
\def\approxprop{\mathpalette\app\relax}

\title[\HI~\& environment of IllustrisTNG galaxies]{Atomic hydrogen in IllustrisTNG galaxies: the impact of environment parallelled with local 21-cm surveys}
\author[A.~R.~H.~Stevens et al.]{Adam R.~H.~Stevens,$^{1,2}$\thanks{E-mail: adam.stevens@uwa.edu.au} Benedikt Diemer,$^3$ Claudia del P.~Lagos,$^{1,2}$ Dylan Nelson,$^4$
\newauthor Annalisa Pillepich,$^5$ Toby Brown,$^6$ Barbara Catinella,$^{1,2}$ Lars Hernquist,$^3$
\newauthor Rainer Weinberger,$^3$ Mark Vogelsberger$^7$ and Federico Marinacci$^3$\\
$^1$International Centre for Radio Astronomy Research, The University of Western Australia, Crawley, WA 6009, Australia\\
$^2$Australian Research Council Centre of Excellence for All Sky Astrophysics in 3 Dimensions (ASTRO 3D)\\
$^3$Institute for Theory and Computation, Harvard-Smithsonian Center for Astrophysics, Cambridge, MA 02138, USA\\
$^4$Max-Planck-Institut f\"{u}r Astrophysik, D-85741 Garching, Bayern, Germany\\
$^5$Max-Planck-Institut f\"{u}r Astronomie, D-69117 Heidelberg, Baden-W\"{u}rttemburg, Germany\\
$^6$Department of Physics and Astronomy, McMaster University, Hamilton, ON L8S 4L8, Canada\\
$^7$Department of Physics, Massachusetts Institute of Technology, Cambridge, MA 02139, USA}

\begin{document}

\defcitealias{gk11}{GK11}
\defcitealias{gd14}{GD14}
\defcitealias{k13}{K13}

\pagerange{5334--5354} \pubyear{2019}

\maketitle

\label{firstpage}

\begin{abstract}
We investigate the influence of environment on the cold-gas properties of galaxies at \zo~within the TNG100 cosmological, magnetohydrodynamic simulation, part of the IllustrisTNG suite.  We extend previous post-processing methods for breaking gas cells into their atomic and molecular phases, and build detailed mocks to comprehensively compare to the latest surveys of atomic hydrogen (\HI) in nearby galaxies, namely ALFALFA and xGASS.  We use TNG100 to explore the \HI~content, star formation activity, and angular momentum of satellite galaxies, each as a function of environment, and find that satellites are typically a factor of $\gtrsim$3 poorer in \HI~than centrals of the same stellar mass, with the exact offset depending sensitively on parent halo mass.  Due to the large physical scales on which \HI~measurements are made ($\sim$45--245\,kpc), contributions from gas not bound to the galaxy of interest but in the same line of sight crucially lead to larger \HI~mass measurements in the mocks in many cases, ultimately aligning with observations.  This effect is mass-dependent and naturally greater for satellites than centrals, as satellites are never isolated by definition.  We also show that \HI~stripping in TNG100 satellites is closely accompanied by quenching, in tension with observational data that instead favour that \HI~is preferentially stripped before star formation is reduced. 
\end{abstract}

\begin{keywords}
galaxies: evolution -- galaxies: haloes -- galaxies: interactions -- galaxies: ISM -- galaxies: star formation
\end{keywords}


\section{Introduction}
\label{sec:intro}

It is well established within the literature that a galaxy's evolution is significantly influenced by its environment \citep[see reviews by][]{dressler84,blanton09,benson10}.  Of primary importance is the position of a galaxy relative to the trough of the local gravitational potential well.  Within a \emph{halo}, \emph{central} galaxies nominally sit at the potential minimum, and therefore act as a sink for cosmological gas that is accreted though the halo -- one (of two) key means by which a galaxy grows.  While all are born as centrals, the mutual gravitational attraction of galaxies (and their haloes) leads to their becoming clustered, with the galaxies of lesser gravitational prominence becoming \emph{satellites} when their haloes coalesce. Satellite galaxies experience variable periods of orbit within the halo, before the majority are fated to merge with their corresponding central (the other key process for centrals' growth).  Satellites are therefore subject to gravitational and hydrodynamical interactions with the halo that centrals do not experience.  Comprising a significant minority of all galaxies in the local Universe (tens of per cent -- e.g.~\citealt{sb17}), the study of satellites and how they are affected by their environment is a crucial piece to complete the puzzle of the field of galaxy evolution.

A plethora of observational evidence has shown that the denser an environment a satellite resides in, the more likely its reservoir of atomic hydrogen (\HI) is depleted  \citep{giovanelli85,solanes01,cortese11,catinella13,stark16,brown17} and -- as \HI~is canonically the raw fuel for future star formation -- the less likely it is to actively form stars \citep{peng10,schaefer17}. There are several key environment-driven processes that can remove or reduce the gas content of satellites and subsequently quench them. Three of the most prevalent processes are (i) starvation, i.e.~the inability of satellites to acquire fresh cosmological gas, exacerbated by the removal of a satellite's circumgalactic medium through hydrodynamical interaction with the intrahalo medium \citep*[][]{larson80};%
\footnote{The latter aspect of starvation is physically a ram pressure, just applied to hot gas rather than cold gas.  Where we refer to `ram-pressure stripping' in this paper, we mean of the ISM.}
(ii) ram-pressure stripping, i.e.~the direct removal of a satellite's interstellar medium (ISM) through hydrodynamical interaction with the intrahalo medium \citep[][]{gunn72}; and (iii) tidal stripping, resulting from variable gravitational interaction across a satellite with the halo or another nearby galaxy \citep[][]{moore99}.  While \emph{direct} observation of starvation is inherently difficult, gaseous tidal stripping has been observed \citep[e.g.][]{marziani03}, and there is an ever-growing number of direct observations of ram-pressure stripping \citep[e.g.][]{chung09,fumagalli14,poggianti17,boselli18}.

Cosmological, hydrodynamic simulations are excellent numerical experiments for developing our understanding of the effect of environment on galaxy evolution.  Because the processes that govern gas stripping are purely hydrodynamical and/or gravitational in nature, the loss of gas in satellite galaxies should occur naturally within these simulations.  That is, \emph{in principle}, these processes are resolved, and not sub-grid like other physical aspects, such as star formation.  This contrasts to semi-analytic models of galaxy formation \citep[see reviews by][]{baugh06,somerville15}, where \emph{all} astrophysical phenomena (including ram-pressure stripping, starvation, and tidal interactions) must be individually, explicitly prescribed.  The trade-off is that this allows for an exploration of the impact each phenomenon has on galaxies by turning each one on and off \citep[e.g.][]{sb17}.  Also contrasting is the ability for semi-analytic models to self-consistently couple gas phases with galaxy evolution, while cosmological, hydrodynamic simulations often do not track the multi-phase nature of gas in detail \citep[for a review of the challenges for next-generation hydrodynamic simulations, see][]{naab17}, and therefore typically must be post-processed to provide information on the likes of \HI~\citep*[but see e.g.][]{dave16}.

Despite some success, to date, no simulation or model has been shown to reproduce \emph{all} the observed effects of gas stripping.  For example, \citet{marasco16} have shown how satellite galaxies in the EAGLE%
\footnote{Evolution and Assembly of GaLaxies and their Environments \citep{schaye15,crain15}.}
hydrodynamic simulations follow the observed trend of \HI~content decreasing with increasing parent halo mass (i.e.~denser environments).  However, at nominal resolution, \HI~masses dropped too rapidly at high parent halo masses, almost making \HI~stripping an all-or-nothing effect (cf.~fig.~6 of \citealt{marasco16} with e.g.~the results of \citealt{brown17}).  In the high-resolution EAGLE runs, this was found to be less of an issue \citep[based on fig.~B1 of][]{marasco16}.  Meanwhile, using the hydrodynamic simulations of \citet{dave13}, \citet{brown17} showed that while the \HI~fractions of satellites in the simulations had a clear dependence on parent halo mass, they were systematically gas-poor.  Similar results were found in that paper with the {\sc galform} semi-analytic model \citep{gonzalez14} -- interestingly, this model has no consideration of ram-pressure stripping of the ISM \citep*[and it is not alone in this regard -- cf.][]{somerville08,henriques15,croton16,hirschmann16,lagos18c}.  And while other semi-analytic models do account for ram-pressure stripping of cold gas \citep*[e.g.][]{stevens16,cora18}, in the case of the {\sc Dark Sage} model, \citet{sb17} show that there is still a systematic discrepancy in the \HI~content of satellites, despite their relative \HI~fractions in parent halo mass bins agreeing reasonably with observations.

A vital step in learning from these simulations and models is in comparing their outcome with observational data.  In doing this, it is critical to measure galaxy properties in a consistent way.  In practice, this means `mock-observing' galaxies in the simulation, using the specifications of the instrument and strategy employed by the survey of interest.  The importance of this step is often overlooked.

In recent years, the completion of large-scale, blind \HI~surveys and deep, targeted \HI~surveys have provided \HI~data for tens of thousands of galaxies in the nearby Universe. The Arecibo Legacy Fast ALFA\footnote{Arecibo $L$-band Feed Array} survey (ALFALFA) is the largest such blind \HI~survey, detecting $\sim$32\,000 galaxies at $z\!<\!0.06$ \citep{giovanelli05,haynes18}, while the eXtended \emph{GALEX}\footnote{\emph{GALaxy evolution EXpolerer}} Arecibo SDSS Survey \citep[xGASS;][]{catinella10,catinella18} is the largest deep, gas-fraction-limited census of \HI~in $\sim$1200 nearby galaxies ($z\!<\!0.05$). The overlap these surveys have with the Sloan Digital Sky Survey (SDSS) means there are accompanying data regarding the stellar mass, star formation rates (SFRs), and host halo masses of these systems.  It is therefore possible to assemble large datasets to look at variations in \HI~content as function of each of these key environment and galaxy properties with statistical significance.

In this paper, we use the main 100-cMpc run from the IllustrisTNG\footnote{Illustris: The Next Generation -- \url{http://www.tng-project.org/}} suite of cosmological, magnetohydrodynamic simulations to study the \HI~properties of galaxies at \zo, focussing on the influence of environment.  We expand upon the methods of \citet{lagos15b} and \citet{diemer18} for breaking the gas cells in the simulations into their atomic and molecular content in post-processing.  Throughout, we make close comparisons to data from the xGASS and ALFALFA surveys.  To do this, we build detailed mocks of the simulation for each survey that replicate many of their design features.  We use these comparisons to demonstrate the power of IllustrisTNG in reproducing the main \HI~scaling relations, and to identify areas in which the simulation is in tension with observations.  The latter is key to further advance our theoretical understanding of galaxy formation and evolution.  Our work complements the likes of \citet{yun18}, who showed that ram-pressure stripping in IllustrisTNG leads to the formation of jellyfish galaxies, in line with observations \citep[e.g.][]{fumagalli14,poggianti17}.

This paper is structured as follows.  In Section \ref{sec:sim}, we describe IllustrisTNG's specifications and our method for the \HI/\Htwo~breakdown of the simulation's gas.  We provide further details on the latter in Appendix \ref{app:equations}.  In Section \ref{sec:obs}, we summarise the xGASS and ALFALFA surveys, and describe our method for mock-observing the simulations for comparison.  Our results begin in Section \ref{sec:comparison}, where we show how the neutral and \HI~content of \zo~centrals and satellites vary with stellar mass in the simulation, comparing closely to the surveys.  Our key results are then presented and discussed in Section \ref{sec:SatEnv}.  We focus on the \HI~content of satellite galaxies in bins of parent halo mass (which we use as a measure of galaxy environment), highlighting the effect of ram-pressure stripping at fixed mass, fixed specific star formation rate (sSFR), and fixed global disc stability (which depends directly on specific angular momentum).  We investigate the relative rate at which \HI~is lost compared to star formation being shut down, finding potential tension between the simulations and survey results.  Discussion of our results is added in Section \ref{sec:conc}, where the paper is also summarised.  We offer some remarks regarding numerical convergence in Appendix \ref{app:res}.


\section{Simulation overview and post-processing method}
\label{sec:sim}

The IllustrisTNG simulations (hereafter `TNG') comprise a suite of magnetohydrodynamic, cosmological simulations for a range of volumes and resolutions.  Each of these follows the standard $\Lambda$CDM cosmological model, with parameters based on the \citet{planck16} results: $\Omega_m \! = \! 0.3089$, $\Omega_{\Lambda} \! = \! 0.6911$, $\Omega_b \! = \! 0.0486$, $h \! = \! 0.6774$, $\sigma_8 \! = \! 0.8159$, and $n_s \! = \! 0.9667$.  The simulations were run with the {\sc arepo} code \citep{springel10}, which discretises gas elements within a moving Voronoi mesh, in order to solve the equations of magnetohydrodynamics.  Poisson's equation for gravity is solved via the tree-particle-mesh method \citep[introduced by][]{xu95}.

Building from the original Illustris simulation and its methods \citep{vogelsberger13,vogelsberger14a,vogelsberger14b,genel14,torrey14}, TNG includes a range of sub-grid models to accommodate relevant astrophysical processes in the formation and evolution of galaxies.  These are described in detail in \citet{pillepich18a} and \citet{weinberger17}.  In short, the simulations account for gas cooling, star formation, growth of massive black holes, and feedback from both stars and active galactic nuclei.

For this work, we use the highest-resolution version of the TNG100 simulation, which has been presented in a series of recent papers \citep{pillepich18b,nelson18,springel18,marinacci18,naiman18}.  TNG100 employs a periodic box of length $75\,h^{-1} \! \simeq \! 100\,{\rm cMpc}$, containing $1820^3$ dark-matter particles of mass $7.5 \! \times \! 10^6\,{\rm M}_{\odot}$, and $1820^3$ initial gas cells.  Stellar particles formed from the gas have a typical mass of $1.4 \! \times \! 10^6\,{\rm M}_{\odot}$, in line with the typical gas cell mass.  Further details about the simulation suite can be found in table 1 of \citet{pillepich18b}.

Parameters in TNG's subgrid prescriptions were manually calibrated to a set of observational constraints using a series of simulations with $25\,h^{-1}\,{\rm cMpc}$ box lengths with $2\!\times\!512^3$ initial resolution elements \citep{pillepich18a}.  Primary constraints included the cosmic star formation rate density history, the \zo~galaxy stellar mass function, and the \zo~stellar--halo mass relation.  Secondary constraints (i.e.~those that were met unless satisfying primary constraints prevented it) included the black hole--bulge mass relation, the gas fraction of haloes within $R_{\rm 500c}$, and the stellar size--mass relation (all at $z\!\simeq\!0$).  We note that almost all of these are exclusive to the stellar content of galaxies -- the exception is the halo gas fractions, which are based on X-ray data.  All results from TNG relating to galaxies' cold-gas content (i.e. \HI~and~\Htwo) are therefore \emph{predictive}.

To be consistent with TNG, where relevant, all results assume $h\!=\!0.6774$ and a \citet{chabrier03} initial mass function throughout this paper.


\subsection{Galaxy finding and sample selection}
\label{ssec:finding}
Gravitationally bound structures in the simulation were found with the {\sc subfind} algorithm \citep{springel01,dolag09}.  Haloes are first identified as particles connected by friends-of-friends (FoF) groups with a linking length of 0.2 times the mean interparticle distance.  Subhaloes are composed of particles/cells that are gravitationally bound to specific substructural over-densities within the halo.  All haloes have at least one subhalo, with the most massive subhalo assigned the `central', and the remainder classed as `satellites'.  It is important to note that the method for assigning centrals and satellites for observational surveys with group finders is quite different (discussed in Sections \ref{sec:obs} and \ref{sec:conc}).

When presenting `inherent' galaxy properties in this paper, we take the particles/cells associated with a subhalo and eliminate any outside a spherical aperture, following the method of \citet{stevens14}.  The idea is to separate what is `part of the galaxy' from what is `part of the rest of the (sub)halo' (i.e.~intrahalo stars and the circumgalactic medium).  In brief, a spherically averaged one-dimensional cumulative mass profile of the stars + neutral gas%
\footnote{\citet{stevens14} originally used `cold' gas, defined by a temperature threshold, rather than neutral gas.  The net intended effect is the same.}
of each subhalo is first built (we describe the neutral gas calculation in Section \ref{ssec:neutral}).  The aperture radius is then set to the radius at which the slope of the mass profile can be described as constant (i.e.~where it is approximately isothermal) within some tolerance.  This has an upper limit of $R_{\rm 200c}$ of the subhalo.  For low-mass systems, this limit is often reached.  Otherwise the resulting radius typically varies between 0.2--0.5$R_{\rm 200c}$.  Compared to using all particles/cells bound to the subhalo, the use of this aperture leads to a visible but subtle and ultimately insignificant difference for inherent gas fractions in the plots we present in this paper.

In addition to exclusively focussing on \zo, we study galaxies in TNG100 with stellar masses $m_* \! \geq \! 10^9\,{\rm M}_{\odot}$.  This cut means we cover the same mass range as the observations we compare against in Section \ref{sec:comparison} and that all the galaxies are sufficiently resolved in the simulations ($\gtrsim\!700$ stellar particles each).  We also exclude any subhaloes whose dark matter accounts for less than 5\% of their total mass; these objects identified by the substructure finder are not actually subhaloes of cosmological origin \citep[for in-depth discussions on subhalo and galaxy finders, see][]{knebe13,canas18}.


\subsection{Neutral fraction of gas cells}
\label{ssec:neutral}

Before considering the atomic or molecular content of gas cells, one first must determine the fraction of gas within a cell that is neutral versus ionized.  For non-star-forming cells, this is trivial from a post-processing stand-point, as the output of TNG already provides a neutral fraction for each cell.  These fractions were calculated in the simulation following the photo-ionizing rate from the ultraviolet background (UVB) of \citet{fg09},%
\footnote{\label{foot:fg09}The December 2011 update of these tables was used, which are publicly available: \url{http://galaxies.northwestern.edu/uvb/}.}
modified to account for self-shielding of ionizing photons \citep[equation A1 of][]{rahmati13a}, and considering cell cooling rates.

For star-forming cells, the internal neutral fraction calculation is inconsistent with the breakdown of gas into its subgrid `ambient hot' and `cold cloud' components in \citet{springel03}, which is also used in TNG.  We therefore instead assume that the `cold cloud' component is entirely neutral, while the remainder is ionized \citep[consistent with][]{marinacci17}.  We provide relevant equations in Appendix \ref{app:neutral}.


\subsection{The atomic- and molecular-phase breakdown of neutral hydrogen}
\label{ssec:hih2}

We recently presented a series of methods for deriving the \HI~and H$_2$ content of simulated galaxies, as applied to TNG \citep{diemer18}, which expanded beyond what had previously been implemented for TNG \citep{villa18}, EAGLE \citep{lagos15b}, and the Auriga simulations \citep{marinacci17}.  For this paper, we employ a subset of those prescriptions, specifically the volumetric versions of those based on \citet[][hereafter GK11]{gk11}, \citet[][hereafter K13]{k13}, and \citet[][hereafter GD14]{gd14}.  For the sake of completeness, we include all equations used in the prescriptions in Appendix \ref{app:molecular}.   We have made minor modifications to the precise methodology of \citet{diemer18} for this paper, translating into differences of integrated galaxy H$_2$ masses of $\lesssim$10\% for all prescriptions (less for \HI). The code for our calculations presented in this paper (inclusive of Section \ref{ssec:neutral} and Appendix \ref{app:equations}) is publicly available and written in {\sc python}.%
\footnote{See the {\tt HI\_H2\_masses()} and {\tt neutralFraction\_SFcells\_SH03()} functions in \url{https://github.com/arhstevens/Dirty-AstroPy/blob/master/galprops/galcalc.py}.  Note that the actual UV field calculation we invoked \citep[identical to][]{diemer18} is not included, although alternative, approximate methods are provided.}
As we will show, any differences in our results from these three prescriptions are almost entirely negligible.  Were we to explore a wider range of prescriptions, the scatter in our results would increase \citep[cf.~the results from][]{diemer18}.

The most important aspect of modelling the \HI/H$_2$ fraction of cells is the strength of the ultraviolet field, specifically that in the Lyman-Werner band, where photons can dissociate molecules \citep[e.g.][]{draine11}.  In what follows, we denote $U_{\rm MW}$ as the strength of the UV field in units of the measured field in the local neighbourhood of the Milky Way, specifically normalising by the flux at 1000\,\AA~in the \citet{draine78} spectrum (in practice, this is equivalent to integrating over the Lyman-Werner band -- see fig.~9 of \citealt{diemer18}).  We assume that star-forming cells emit UV with an intensity proportional to their star formation rate, whose flux at 1000\,\AA~is based on that of the equilibrium spectrum of a continuously forming population of stars, calculated with {\sc Starburst99} \citep{leitherer99}.  90\% of the UV is assumed to be absorbed by the cell of origin, while the rest propagates through neighbouring cells, which we treat as an optically thin medium.  The propagation allows us to model the UV strength in non-star-forming cells, rather than assuming it to be negligible.  We further enforce a minimum value of $U_{\rm MW}$ for all cells from the \citet{fg09} UVB.  Our UV treatment is originally presented in \citet{diemer18}, and we refer the reader to that paper for further details (including associated uncertainties).  

One difference in our implementation for this paper is that, where \citet{diemer18} processed subhaloes individually, we process entire FoF groups at a time.  We loop over all subhaloes within the group, where, for each, we centre on them and follow the details in appendix A2 of \citet{diemer18}.  The UV flux through a given cell in a group is taken as its maximum from all loops.  Allowing for all galaxies to contribute to the UV flux for each cell is important for how diffuse gas factors into our integrated \HI~mass measurements (see Section \ref{sec:obs}); i.e.~if we only used the central galaxy to inform the UV flux throughout of the intrahalo medium, we would underestimate it, especially in regions near satellites.

For all the prescriptions that we implement (i.e.~those in Appendix \ref{app:molecular}), one needs a way of converting between three-dimensional cell densities, $\rho$ (inherent to the simulation), and two-dimensional surface densities, $\Sigma$ (which the models for the \HI/H$_2$ fraction are based on).  For this, we use the Jeans length of the cell, $\lambda_J$.  Assuming the ideal gas law, this gives
\begin{equation}
\Sigma = \rho\, \lambda_J = \sqrt{\frac{\gamma\, k_B\, T\, \rho}{\mu\, m_p\, G}} = \sqrt{\frac{\gamma\, (\gamma-1)\, u\, \rho}{G}}\,,
\label{eq:jeans}
\end{equation}
where $k_B$ and $G$ are the Boltzmann and gravitational constants, respectively, $\gamma$ is the ratio of specific heats, and we have assumed all gas is thermalised \citep[cf.][]{schaye08}.  The neutral-hydrogen surface density of gas is hence
\begin{equation}
\Sigma_{\rm H\,{\LARGE{\textsc i}}+H_2} = X\, f_n\, \Sigma\,.
\end{equation}
While \citet{diemer18} showed that the Jeans approximation (i.e.~Equation \ref{eq:jeans}) is not particularly accurate in detail, similar results \emph{on average} were produced when the \HI/\Htwo~fraction was modelled in projection instead (but for individual galaxies, these two methods gave non-negligible differences).

The precise value of $\gamma$ (and $\mu$) depends on the atomic-to-molecular ratio.  To solve Equation (\ref{eq:jeans}) for the \HI/H$_2$ fraction would thus require already knowing the answer.  To resolve this, every prescription is applied iteratively, until self-consistent values for $\gamma$ and the molecular fraction are obtained.  We approximate
\begin{equation}
\gamma = \frac{5}{3}(1-f_{\rm mol}) + \frac{7}{5}f_{\rm mol}\,.
\end{equation}
Here,
\begin{subequations}
\begin{equation}
f_{\rm mol} \equiv \frac{X\, f_n\, f_{\rm H_2}}{1-Z}\,,
\end{equation}
\begin{equation}
f_{\rm H_2} \equiv \frac{\rho_{\rm H_2}}{\rho_{\rm H\,{\LARGE{\textsc i}}} + \rho_{\rm H_2}} = \frac{\Sigma_{\rm H_2}}{\Sigma_{\rm H\,{\LARGE{\textsc i}}+H_2}}\,.
\end{equation}
\end{subequations}
Our iterative method here is a new feature relative to \citet{diemer18}, which theoretically offers \emph{modestly} more accurate \HI~and \Htwo~cell masses.

Another common number for all prescriptions is the dust-to-gas ratio relative to the Milky Way.  This ratio is assumed to be equal to the metallicity ratio of a gas cell to the Milky Way, and is hence given as
\begin{equation}
D_{\rm MW} = Z / 0.0127
\end{equation}
\citep[as in][]{lagos15b}.  The physical role of dust is important, as not only do molecules form on dust grains, but dust also helps shield molecules from dissociating photons.  This aspect of our model has plenty of room for development \citep*[see, e.g., the dust models incorporated into {\sc arepo} simulations by][]{mckinnon16,mckinnon17}

\begin{figure*}
\centering
\includegraphics[width=\textwidth]{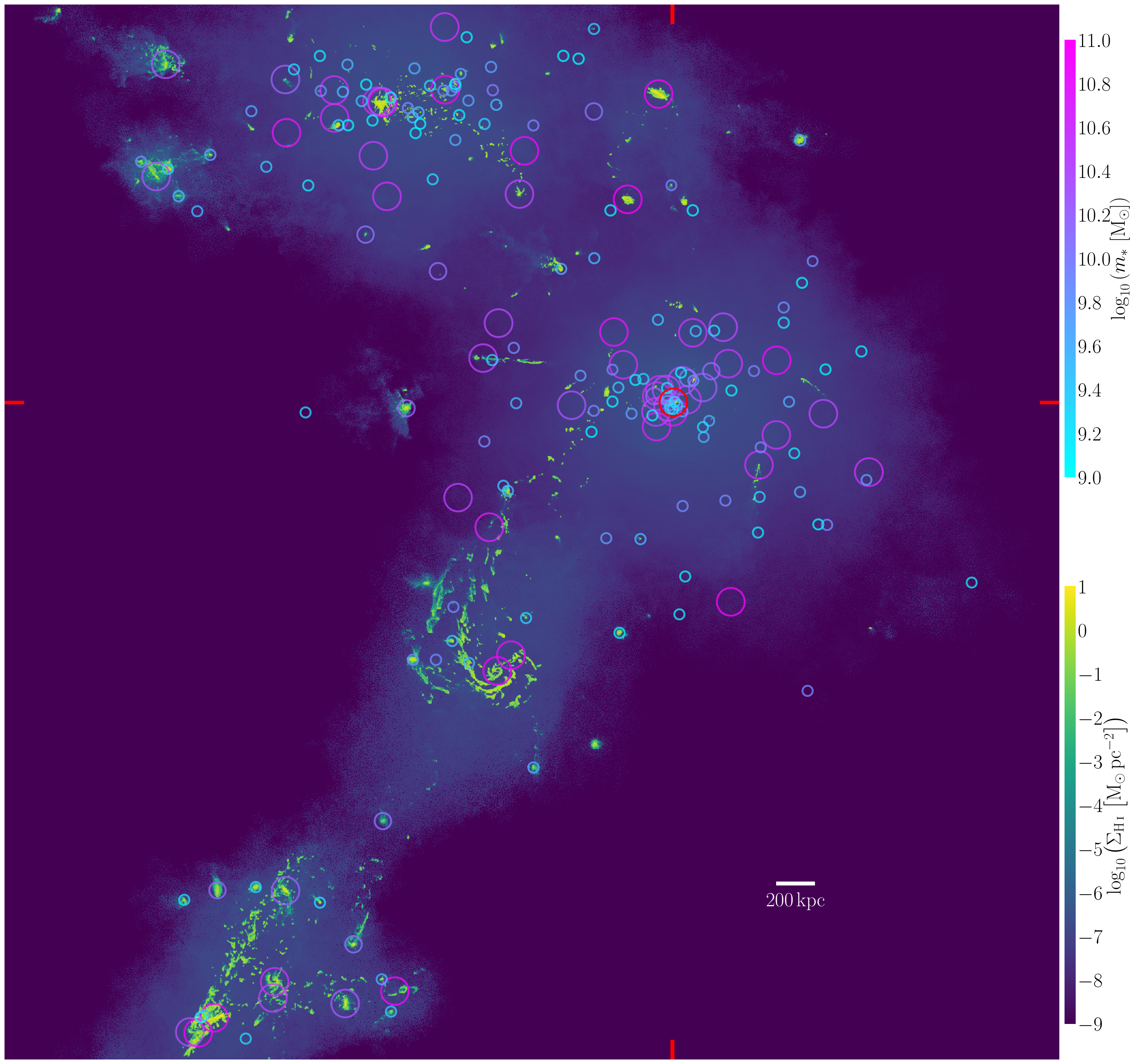}
\caption{Projected \HI~column density from all gas cells in Group 3 (with $M_{\rm 200c}\!=\!1.7\!\times\!10^{14}\,{\rm M}_{\odot}$) of TNG100 at \zo.  The image is $6\!\times\!6$\,Mpc.  Circles indicate the physical beam size (circle diameter = full width at half maximum) used to measure \HI~masses in this cluster for the xGASS mock (Section \ref{ssec:xgass}), and are coloured by the stellar mass of the galaxy they are centred on.  The central galaxy of the cluster can be identified by the deeper red circle, located where the red edge marks would meet.}
\label{fig:image}
\end{figure*}

To showcase our method, we present an image of a group in TNG100 in \HI~(using the \citetalias{gd14} method) in Fig.~\ref{fig:image}.  We also show the scale on which the integrated \HI~mass of each galaxy in this group is measured for the xGASS mock survey measurements, which we outline in the next section.

\subsection{Disc properties}
\label{ssec:disc}

In Sections \ref{ssec:phase} \& \ref{ssec:qf} of our results, we refer to disc masses and radii of TNG100 galaxies.  To calculate these, we first identify stellar particles and gas cells whose motions are close to a circular orbit about the galaxy centre of mass.  Specifically, we apply the criteria of \citet[][see their equations 3 \& 4 -- when applying this for stellar particles, we assume the term for internal energy is zero]{mitchell18}.  We then find the smallest cylinder that encompasses all these particles/cells, ensuring its orientation matches that of the galaxy's angular-momentum vector.  We denote the radius of this cylinder as $r_{\rm disc}$.  Our disc mass for a galaxy therefore includes all particles/cells in this cylinder of the relevant species, even if they are not rotationally supported.


\section{Observational data and comparison method}
\label{sec:obs}

Throughout this paper, we compare results of TNG100 with recently published observational data concerning the \HI~content of galaxies in the local Universe.  These data originate from two surveys: ALFALFA \citep{giovanelli05,haynes11} and xGASS \citep{catinella10,catinella18}.  In this section we describe the data themselves, and how we have measured galaxies' properties in TNG to compare against the data in the fairest manner, effectively `mock-observing' the galaxies in their \HI~content.


\subsection{The xGASS representative sample}
\label{ssec:xgass}

The \emph{GALEX} Arecibo SDSS Survey \citep[GASS;][]{catinella10} observed the 21-cm \HI~line emission of galaxies at $ z \! \in \! [0.025,0.05]$, probing down to stellar masses of $10^{10}\,{\rm M}_{\odot}$.  Each galaxy was observed for sufficiently long such that an \HI~detection was effectively guaranteed should galaxies have $m_{\rm H\,{\LARGE{\textsc i}}} \! \gtrsim \! {\rm max}\!\left[10^{8.7}\,{\rm M}_{\odot},\,0.015\,m_* \right]$.  Using a similar observing strategy, the follow-up GASS-low survey \citep{catinella18} then extended observations down to stellar masses of $10^{9}\,{\rm M}_{\odot}$ in the redshift range $0.01 \! \leq \! z \! \leq \! 0.02$, with detections in \HI~for $m_{\rm H\,{\LARGE{\textsc i}}} \! \gtrsim \! {\rm max}\!\left[10^{8}\,{\rm M}_{\odot},\,0.02\,m_* \right]$.  Both surveys have relatively flat redshift distributions, although there is a weak mass dependence on the redshifts of galaxies in GASS.  GASS and GASS-low overlap in the mass range $\log_{10}\left(m_* / {\rm M}_{\odot} \right) \! \in (10, 10.23)$.  The `xGASS representative sample' is comprised of data from GASS, GASS-low, and includes detections from ALFALFA and the \citet{springob05} catalogue.  These data are publicly available,%
\footnote{\url{http://xgass.icrar.org/data.html}}
and we hereafter simply refer to this sample as `xGASS'.

Because all the xGASS galaxies are observed with Arecibo, when comparing \HI~properties of TNG galaxies, we need to mock-observe them in a manner that is consistent with how Arecibo would see them.  Integral to this is that we place each galaxy at a distance to convert the angular beam size of Arecibo to a physical beam size.  We assume that all galaxies in the same FoF group in the simulation are observed together (modulo the caveat to follow), and that each FoF group is independent in its mock-observed distance.  We build probability distribution functions (PDFs) of redshifts of xGASS central galaxies in bins of stellar mass, and draw from these PDFs to assign each central galaxy in TNG100 a cosmological redshift that can be directly translated into an angular-diameter distance.  Nominally, the distance of each satellite is taken as that of the central plus the inherent $z$-direction separation (not to be confused with redshift) of the satellite from the central.  In truth, it is only this simple if the satellite and central fall in the same of three mass ranges: $m_* \! \leq \! 10^{10}\,{\rm M}_{\odot}$, $m_* / {\rm M}_{\odot} \! \in (10^{10}, 10^{10.23})$, and $m_* \! \geq \! 10^{10.23}\,{\rm M}_{\odot}$, corresponding to those exclusively in GASS-low, those in the overlap range of GASS and GASS-low, and those exclusively in GASS, respectively.  When a central has $m_* \! \geq \! 10^{10.23}\,{\rm M}_{\odot}$, we effectively assign 3 redshifts for its group.  Secondary and tertiary redshifts are drawn from the observed PDFs of xGASS \emph{satellite} redshifts in the overlap mass range and below, respectively.  Satellites in that group that fall in either of these mass ranges are then mock-observed at that respective redshift.  Similarly, if the central's mass places it in the overlap range, we assign a secondary redshift for the group that satellites with $m_* \! \leq \! 10^{10}\,{\rm M}_{\odot}$ use.  If the central lies in the lowest mass range, then all galaxies in that halo use the same redshift.  

\begin{figure}
\centering
\includegraphics[width=\textwidth]{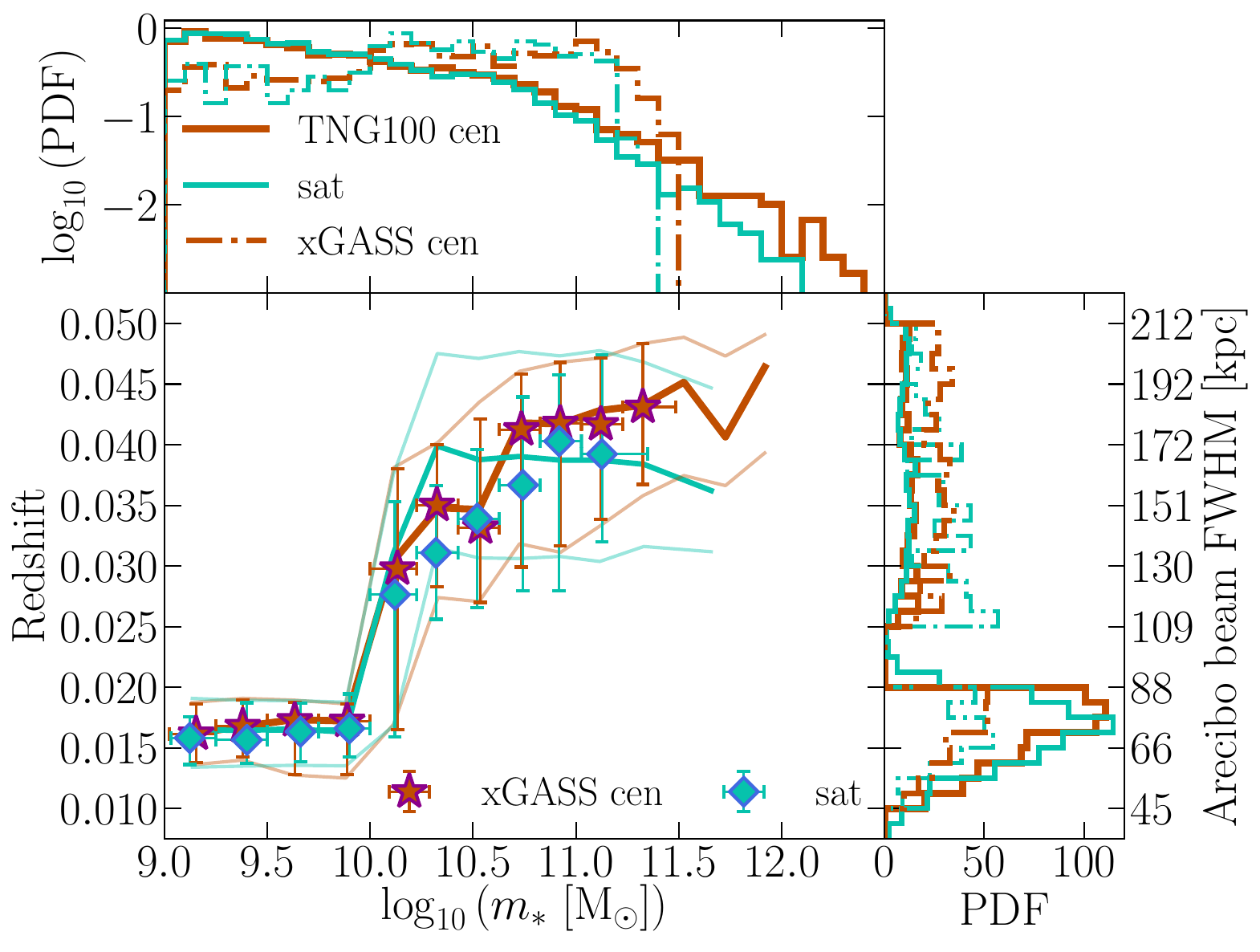}
\includegraphics[width=\textwidth]{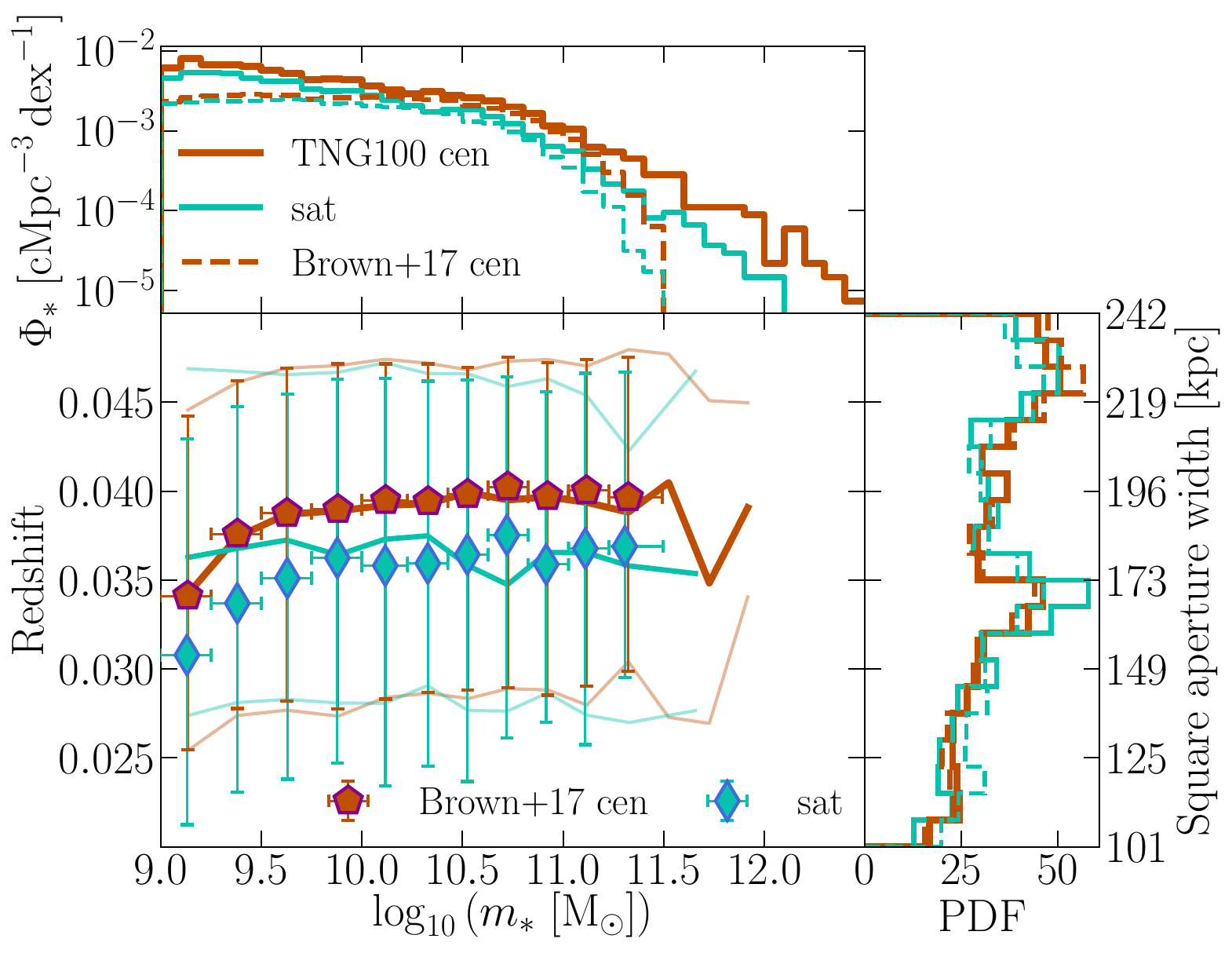}
\caption{Distributions of TNG100 galaxies (separated into centrals and satellites) in redshift and stellar mass for xGASS (top) and ALFALFA (bottom) mocks, compared to the actual distributions from those surveys.  Points and thick lines in the main panels are medians, whereas thin lines and vertical error bars give the 16$^{\rm th}$ and $84^{\rm th}$ percentiles.  Horizontal error bars cover the bin width. Centrals in the mocks are assumed to have no peculiar velocity, while the $z$-direction velocity of satellites relative to their central in the simulation is taken as their peculiar velocity, which is incorporated into their mock redshift.  For each marked redshift, the physical scale on which \HI~masses are measured is indicated for each mock.   Smaller panels give the one-dimensional projected probability distribution functions along each axis. xGASS has a flat stellar-mass distribution by design.  For the lower stellar-mass distribution plot, we have normalised by the respective comoving volumes of the simulation and survey, thereby providing stellar mass functions for the satellites and centrals (the \citealt{brown17} sample is volume-limited).}
\label{fig:zdists}
\end{figure}

We stress that each galaxy in the simulation only contributes to the final xGASS mock once.  By construction, our method ensures we recover a sample of central galaxies in TNG whose distribution in the two-dimensional redshift--stellar mass plane matches xGASS, as shown in Fig.~\ref{fig:zdists}.  By default, it does not guarantee the same distribution for satellites will be precisely recovered.  To mitigate this, we built several thousand mocks following the above method, then chose the one whose median redshift as a function of stellar mass for satellites had the lowest $\chi^2$ when compared with xGASS.  Note that the redshift of each satellite in the mock also incorporates its relative velocity to its central (effectively its peculiar velocity).

The nominal beam diameter for Arecibo is 3.5\,arcmin.  We interpret this to mean the response decreases as a function of beam radius like a Gaussian, where the full width at half maximum (FWHM) of that Gaussian is 3.5\,arcmin (equivalent to a standard deviation of $\sim$1.5\,arcmin).   In principle, any \HI~in this column with a velocity close to that of the galaxy of interest will contribute to the measured \HI~mass of that galaxy.  As we process each TNG halo, we loop over all galaxies in the halo, identify gas cells whose velocity component in the line-of-sight direction (inherent $z$-direction) is within a mass-dependent threshold, and assign each of those cells a weight based on their $xy$-planar distance from the galaxy's centre of mass, according to the Gaussian beam response.  A weighted sum of those cells' \HI~masses gives the galaxy's mock-observed \HI~mass.  The velocity threshold, $v_{\rm thresh}$, approximates the typical projected rotational velocity of galaxies at the relevant mass, accounting for the inherent inclination of the galaxy, $i$:
\begin{equation}
\label{eq:vthresh}
\log_{10}\left(\frac{v_{\rm thresh} \csc{i}}{\rm km\,s^{-1}} \right) = 0.2 \left[1+ \log_{10}\left(\frac{m_*}{\rm M_{\odot}} \right) \right]\,.
\end{equation}
This is effectively just a Tully--Fisher relation \citep{tully77}, where $v_{\rm thresh}$ proxies the half-width of a galaxy's 21-cm line (if present).  We enforce a minimum $v_{\rm thresh}$ of 35 km\,s$^{-1}$, which is a few times the typical \HI~velocity dispersion of discs \citep[cf.][]{tamburro09}; this is only relevant for galaxies close to being face-on.  Equation (\ref{eq:vthresh}) is a rounded form of the Tully--Fisher relation from \citet{reyes11}, after modifying for a \citet{chabrier03} IMF, assumed value of $h$, and use of MPA-JHU SDSS DR7%
\footnote{Max Planck institute for Astrophysics--John Hopkins University~~Sloan Digital Sky Survey~~Data Release 7~~catalogue, available at \url{https://wwwmpa.mpa-garching.mpg.de/SDSS/DR7/}, with improved stellar masses from \url{http://home.strw.leidenuniv.nl/\~jarle/SDSS/}.}
stellar masses (as opposed to those of \citealt{bell03}), which is where stellar masses for xGASS originate.  Stellar mass estimates from the MPA-JHU catalogue are computed by multiplying the dust-corrected luminosity by the predicted mass-to-light ratio from the best-fitting model of the spectral energy distribution (SED); in principle, these should be total stellar masses, and not just restricted to the SDSS fibre aperture.

Although representative in terms of \HI~content, xGASS is not complete; the two volume-limited surveys enforced approximately flat distributions of galaxies in terms of stellar mass.  With a similar but modified method to that of \citet{catinella18}, we therefore weight each xGASS galaxy when calculating medians and percentiles, according to the expected frequency of galaxies at that mass versus what is in the sample.  To calculate these weights, we use finer bins of 0.1\,dex (half the width of the typical bin for which we present results) to histogram galaxy counts for centrals and satellites separately for both xGASS and TNG100.  The weight assigned to the galaxies in each finer bin is then the normalised ratio of counts in TNG100 to xGASS.  By using TNG100 galaxy counts rather than a stellar mass function from another survey to set the weights, we ensure the fairest comparison to TNG100.

xGASS galaxies are tagged as centrals and satellites according to the modified \citet{yang07} catalogue, where cases of `galaxy shredding' and false pairs have been resolved \citep{janowiecki17}.  The \citet{yang07} catalogue is based on the friends-of-friends (FoF) group-finding algorithm of \citet{yang05}, applied to the seventh data release (DR7) of SDSS \citep{abazajian09}, where the galaxy with the greatest stellar mass in the group is assigned as the central (the rest are satellites).  To mock an uncertainty on stellar masses in TNG that roughly matches SDSS, we add to $\log_{10}\left(m_*\right)$ of TNG galaxies a random number drawn from a Gaussian of standard deviation 0.08\,dex \citep[as done in, e.g.,][]{knebe18b}.  We further reassign centrals and satellites in the mock, flagging the galaxy with the greatest mock stellar mass as the central, and reassign the rest to be satellites (the inherent central in TNG is the most massive subhalo, which need not necessarily have the greatest \emph{stellar} mass, although it does for $\sim$98\% of cases).

A subsample of the xGASS galaxies have also been observed in carbon monoxide emission as part of the COLD GASS\footnote{CO Legacy Database for GASS} and COLD GASS-low surveys \citep{saintonge11,saintonge17}, collectively referred to as xCOLD GASS, providing \Htwo~masses for these galaxies.  The overlap in xGASS and xCOLD GASS is referred to as xGASS-CO \citep{catinella18}.  In this paper, xCOLD GASS data are only used for establishing the total neutral-hydrogen content of galaxies (Fig.~\ref{fig:NeutralFrac}).  In an accompanying paper (Stevens et al.~in prep.), we will use xCOLD GASS data to more specifically compare the H$_2$ content of TNG galaxies.  In this work, we assume the metallicity-dependent CO-to-\Htwo~conversion function of \citet{accurso17} and the aperture corrections of \citet{saintonge12}.  All xCOLD GASS data were sourced from the publicly available catalogue.%
\footnote{\url{http://www.star.ucl.ac.uk/xCOLDGASS/data.html}}
We subtracted the `helium contribution' \citep{saintonge17} from these masses.

We compare TNG100 results to xGASS-CO and the full xGASS representative sample in Figs \ref{fig:NeutralFrac} \& \ref{fig:HIFrac}, respectively.  These are discussed in Section \ref{sec:comparison}.


\subsection{Stacked ALFALFA \HI~data}

\label{ssec:alfalfa}

The ALFALFA survey \citep[see][]{giovanelli05,haynes18} is a blind \HI~survey that has measured the 21-cm emission of $\sim$7000\,deg$^2$ of sky out to $z\!<\!0.06$.  Within this volume, $\sim$32\,000 galaxies have their \HI~content detected.
Due to the faintness of 21-cm emission and the short integration times of ALFALFA, the majority of the survey's detections are restricted to the gas-rich and, therefore, star-forming regime. To combat this, we use data from \citet{brown17}, who employed the \HI~spectral stacking technique, co-adding both 21-cm detections and non-detections, to obtain average \HI~scaling relations as a function of key global galaxy properties for a representative sample of galaxies. This sample is selected according to stellar mass -- $m_* \! \in \! \left[10^9,10^{11.5}\right] {\rm M}_{\odot}$ -- from the overlap in volume between SDSS DR7 and the ALFALFA $\alpha.70$ data release \citep{haynes11} between $0.02 \!\leq\! z \!\leq\! 0.05$, totalling $\sim \! (150\,{\rm cMpc})^3$ of the local Universe (more than thrice the comoving volume of TNG100). Redshifts, stellar masses, and SFRs were extracted from the MPA-JHU SDSS DR7 catalogue,
adjusted where necessary for a \citet{chabrier03} initial mass function. For a full description of this technique and the samples used, we refer the reader to \citet{brown15,brown17}. Finally, because we consider \HI~\emph{fractions}, the stacking procedure weights each galaxy's 21-cm spectrum by its stellar mass, thereby returning the mean \HI~fraction, $\langle m_{\rm H\,{\LARGE{\textsc i}}} / m_* \rangle$, for a given stacked galaxy population.  We compare results from TNG100 with these data in Figs \ref{fig:HIFrac} \& \ref{fig:SatEnv}, discussed in Sections \ref{sec:comparison} \& \ref{sec:SatEnv}, respectively.

The 21-cm spectrum for each galaxy was extracted from the `all sky' data cube, using a $4\!\times\!4$ arcmin square aperture around the galaxy's position.  When mock-observing TNG galaxies for comparison to ALFALFA, there are two differences to the method described in Section \ref{ssec:xgass}.  First, we assign each central galaxy with a new, cosmological redshift in the range $[0.02,0.05]$, again matching the survey's distribution of centrals in the two-dimensional redshift--stellar mass plane (see Fig.~\ref{fig:zdists}).  Because the survey was conducted over a single redshift interval, we do not need secondary or tertiary redshifts for galaxies as in Section \ref{ssec:xgass}.  Next, we extract all gas cells in a square aperture (whose edges are parallel to the inherent $x$- and $y$-directions of the simulation, with lengths that translate to 4\,arcmin for the galaxy's mock redshift) that fall in the velocity threshold of Equation (\ref{eq:vthresh}).  No weights are assigned to these cells; their masses are simply summed to give an integrated \HI~mass (the beam response in ALFALFA is already accounted for in the data).
This method means that for galaxies that are close together in both projected distance and line-of-sight velocity, some blending of \HI~emission can take place (this is also true for the xGASS mock). We note that \citet{jones15} estimate this `confusion' rate in ALFALFA to be less than $10\%$, and that any blending must increase the mean gas fraction. An example of confusion can be seen by the overlap of apertures in Fig.~\ref{fig:image} (while these are not the apertures for the ALFALFA mock, they are similar in size).
Our method for calculating \HI~masses for the ALFALFA mock builds detail on that implemented by \citet{marasco16} in comparing the \HI~content of EAGLE galaxies with stacked ALFALFA data.  Specifically, those authors applied a \emph{circular} aperture of \emph{fixed} physical diameter (150\,kpc) with a \emph{fixed} velocity threshold (400\,km\,s$^{-1}$).

\begin{figure}
\centering
\includegraphics[width=\textwidth]{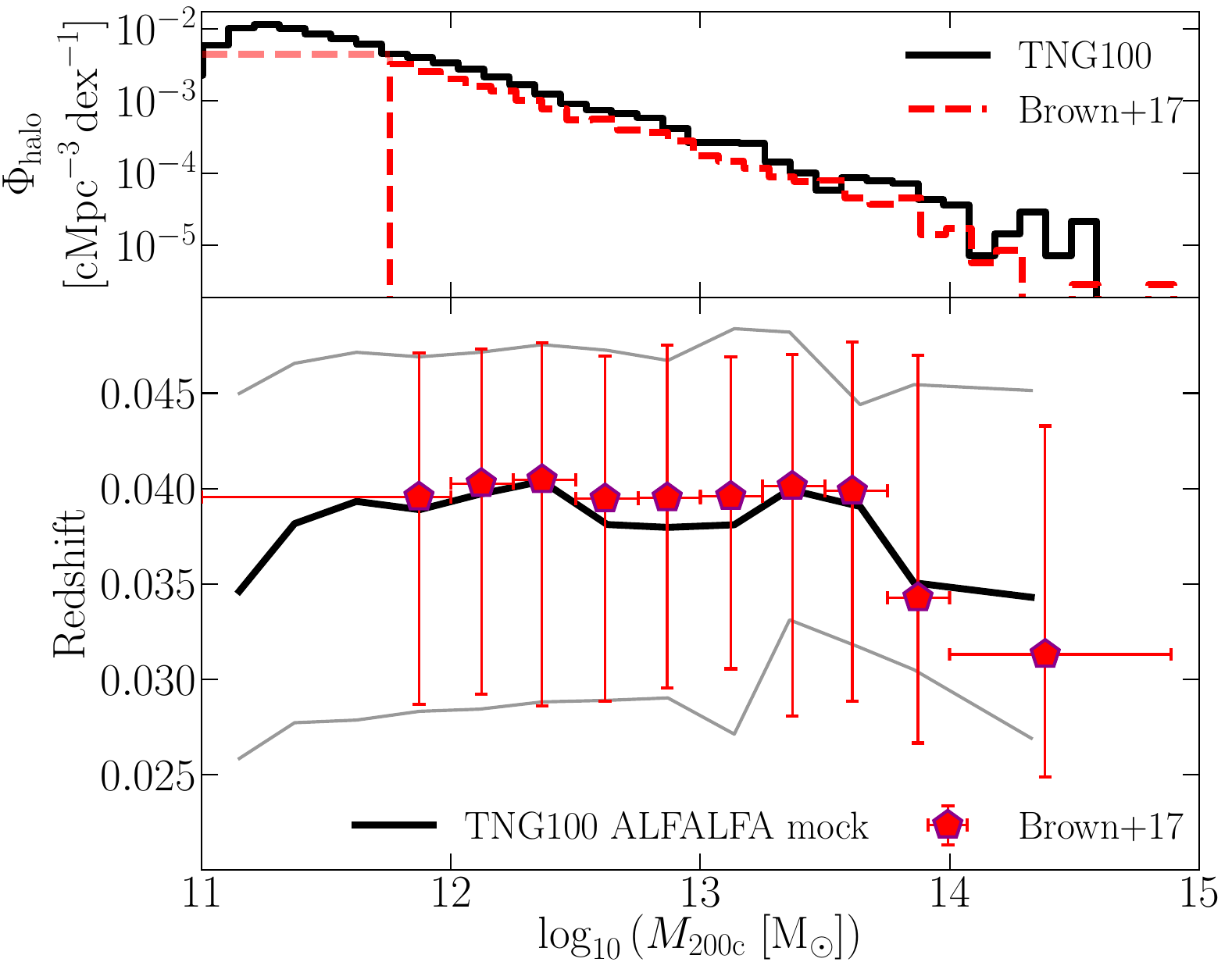}
\caption{Top panel: halo mass functions of TNG100 and the \citet{brown17} ALFALFA sample (the latter is based on the \citealt{yang07} SDSS catalogue).  Bottom panel: distribution of haloes in terms of mass and redshift in the observational sample and ALFALFA mock for TNG100. Points and thick lines are medians, while thin lines and vertical error bars cover the running 16$^{\rm th}$ and $84^{\rm th}$ percentiles.  Horizontal error bars span the full bin width. Haloes masses were not actually assigned in the SDSS catalogue for the lowest-mass bin used here; for the sake of visualising their inclusion, we assume that bin starts at $10^{11}\,{\rm M}_{\odot}$ (a fainter line in the top panel is used to signify this).}
\label{fig:zdist_halo}
\end{figure}

As described in \citet{brown17}, host halo masses for each galaxy in the ALFALFA sample are obtained from the \citet{yang07} `modelB' catalogue.  As per Section \ref{ssec:xgass}, the galaxy with the highest stellar mass in a group is considered the central.  We note that \citet{campbell15} find that 30--40\% of the galaxies classed as satellites by this algorithm are likely actually centrals, while $\sim$20\% of galaxies classed as centrals are likely actually satellites.  In an attempt to mediate this, we not only follow Section \ref{ssec:xgass} in adding uncertainty to stellar masses and reassigning centrals and satellites for the ALFALFA mock for TNG100, but we also reassign host halo masses such that their rank ordering matches that of their total stellar mass, reflecting the abundance matching step used by the group finder for SDSS.  While consideration of these effects does not mimic all the features introduced by the group finder, it does offer a fairer comparison.  We refer the reader to \citet{sb17} for further assessment of the impact that central/satellite impurities can have on model results when compared to the same observational data with the same group catalogue.  Because the halo mass bins we use are so wide ($\sim$1\,dex), these steps have a minimal impact on producing the mock compared to the \HI~measurements.  For completeness, we show and compare the distribution of haloes in our mock in terms of their mass and redshift in Fig.~\ref{fig:zdist_halo}.

Following \citet{brinchmann04}, SFRs for active galaxies in the MPA-JHU SDSS DR7 catalogue are typically measured from H$\alpha$, probing time-scales of tens of Myr.  Passive galaxies, on other hand, usually have SFRs measured from the 4000-\AA~break, probing the average historical SFR of the galaxy over the last $\sim$1\,Gyr.  The inherent SFRs of galaxies in TNG come from the instantaneous SFRs of the galaxy's gas cells, which provide a probability for when a stellar particle should be generated.  For the ALFALFA mock of TNG100, we remeasure these SFRs based on the appropriate time-scales.  First, using inherent SFRs, we break the galaxy population into passive and active following the method of \citet{davies18}, which results in a simple, precise divide at ${\rm sSFR} \! = \! 10^{-11}\,{\rm yr}^{-1}$ for TNG100.  For the active population, we then sum the \emph{birth} mass of stellar particles whose ages are less than 20\,Myr, dividing by this same time-scale.  This provides our resulting mock SFR.  We repeat the same process for stellar-particle ages of 1\,Gyr for the passive population.  This aspect of the mock is relevant for Fig.~\ref{fig:SatEnv}.  Note that SFRs from the MPA-JHU catalogue are aperture-corrected; following \citet{salim07}, corrections had been applied by performing SED fits to the broad-band photometry to derive the SFR for the area of the galaxy not covered by the SDSS fibre.  We therefore account for all stellar particles in the subhalo when calculating mock SFRs for TNG galaxies.  To account for observational uncertainty, when binning galaxies by sSFR for the TNG mocks (i.e.~in Figs \ref{fig:SatEnv} and \ref{fig:ssfr_mstar}), we convolve each SFR by a lognormal distribution of standard deviation 0.29\,dex for the active population, and 0.54\,dex for the passive population \citep[from table 1 of][]{salim07}.

While the \citet{brown17} sample excludes galaxies with $m_* \! > \! 10^{11.5}\,{\rm M}_{\odot}$ to ensure mass completeness, we extend some plots beyond this mass for the TNG100 mocks where stellar mass is on the $x$-axis, as to not unnecessarily disregard information from the simulation.


\section{Cold-gas content of centrals and satellites in the local Universe}
\label{sec:comparison}

\begin{figure*}
\centering
\includegraphics[width=\textwidth]{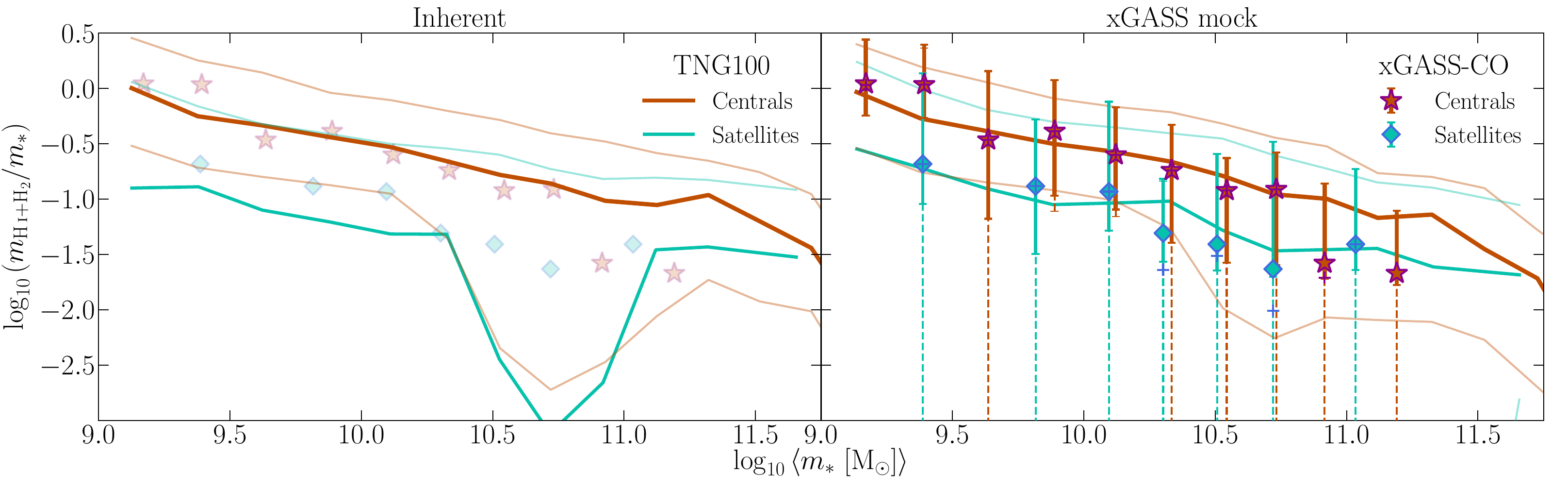}
\caption{Ratio of mass in neutral hydrogen to that in stars for galaxies at \zo, as a function of their stellar mass.  The left panel accounts for all gas cells that are both bound to the subhalo (i.e.~those identified by {\sc subfind}) and within the `BaryMP' radius of \citet{stevens14}.  The right panel follows the mock-observing strategy described in Section \ref{ssec:xgass}, which accounts for the beam response of Arecibo, the redshift distribution of xGASS galaxies, and the SDSS central/satellite definition.  Lines follow the running median and running 16$^{\rm th}$ \& 84$^{\rm th}$ percentiles from TNG100.  Starred and diamond points give the observational medians when non-detections assume their upper limit value (for centrals and satellites, respectively), while crosses give the medians when non-detections are set to zero; where these are not visible, the median value \emph{is} a non-detection.  Error bars cover the 16$^{\rm th}$--84$^{\rm th}$ percentile range; thick, solid error bars assume upper-limit values in the case of non-detections, while the thin, dashed ones set non-detections to zero.  Faded xGASS-CO medians are also included in the left panel -- these should not be directly compared to the inherent TNG100 properties, but are there to help highlight the differences in TNG100 results \emph{between} the two panels.  For both the simulation and observed data, bins of minimum width 0.2\,dex (0.25\,dex for $m_*\!<\!10^{10}\,{\rm M}_{\odot}$) with a minimum of 20 galaxies (relaxed to 16 for xGASS-CO satellites exclusively, given their small total) were used to produce the medians and percentiles.  The plotted $x$-axis position of each bin is the mean stellar mass of galaxies in that bin.  All samples are representative for $m_* \! \geq \! 10^9\,{\rm M}_{\odot}$.}
\label{fig:NeutralFrac}
\end{figure*}

Before specifically addressing the atomic-hydrogen content of galaxies in TNG100, it is important to understand how their total neutral-hydrogen content (i.e.~\HI+\Htwo) compares with observations -- this is, of course, independent from any \HI/\Htwo~prescription.  We present the neutral-hydrogen fraction (where `fraction' here means ratio to stellar mass) of TNG100 galaxies in Fig.~\ref{fig:NeutralFrac}, where we have separated centrals and satellites.  The left panel shows the `true' properties of the simulated galaxies, in that their gas masses have been calculated using cells exclusively associated with the subhalo and within a spherical aperture designed to separate `the galaxy' from `the rest of the subhalo' \citep[see][summarised in Section \ref{ssec:finding} of this work]{stevens14}.  The right panel remeasures TNG100 neutral-hydrogen masses as if they were observed with Arecibo (Section \ref{ssec:xgass}), and thus provides the most meaningful comparison to xGASS-CO.

By and large, the neutral gas fractions of galaxies in TNG100 agree very well with observations.  Satellites in particular trace both the median and 84$^{\rm th}$ percentile of xGASS-CO within $\sim$0.14\,dex on average across the full mass range.  The 16$^{\rm th}$ percentile for TNG100 satellites is off the plot; these galaxies are typically devoid of gas entirely.  This is consistent with observations, in that the galaxies along this percentile are all non-detections in both \HI~and CO.  TNG100 centrals generally trace the observations too, but begin to diverge at $m_* \! \gtrsim \! 10^{10.7}\,{\rm M}_{\odot}$, where they are evidently too gas-rich (the medians for xGASS-CO are $>$0.5\,dex below that of TNG100 in the last two bins).

Mock-observing the TNG100 galaxies washes away a very noticeable feature in the left panel of Fig.~\ref{fig:NeutralFrac}.  The running median for satellites shows a significant ($\sim$1.5\,dex) dip in neutral fraction around $m_* \!\gtrsim\! 10^{10.5}\,{\rm M}_{\odot}$, which persists until $m_* \!\lesssim\! 10^{11}\,{\rm M}_{\odot}$. To some extent, this is also seen in the running 16$^{\rm th}$ percentile for centrals, although that dip is still present in the right panel too.  In fact, the neutral fractions of satellites in general are raised noticeably when mock-observed.  This is because gas in the halo that is not gravitationally bound to the satellites (as determined by {\sc subfind}), but is close to them in both physical and velocity space, contributes towards their measured gas content.  Gas in other galaxies in the same line of sight can also have their mass double-counted or `confused'.  For centrals, not only is the diffuse halo gas already typically associated with them by {\sc subfind} (that gas is still bound to the halo as a whole), but any satellites in the same line of sight will only contribute a relatively small fraction of the measured gas mass \emph{in general} (the central is defined as the most-massive galaxy in the halo).\\

The dip in satellites' \HI~fractions is nevertheless curious, and one should question whether this is a genuine forecast for observations or a feature of the simulation that is not representative of reality.  Let us assess the evidence for each case, considering that from both (i) observations and (ii) simulations:
\begin{enumerate}
\item If the dip were real, it would be reasonable to expect that if equivalent satellites from GASS (i.e. with $m_* \! \simeq \! 10^{10.75}\,{\rm M}_{\odot}$ and median gas fractions for their mass) had their \HI~discs fully resolved (e.g.~with radio interferometry), integrating the mass of the discs would give a much lower measurement than the single-dish Arecibo pointing.  Many studies have compared single-dish and interferometric \HI~mass measurements, and while there is some galaxy-to-galaxy variation, there is no strong evidence of these generally being significantly different \citep*[see][]{kamphuis96,swaters02,walter08,gereb16,gereb18,pingel18}.  However, we could not identify a study that included analogues of those responsible for the dip in TNG100.  Many examples in the literature are at $z \! \lesssim \! 0.002$, meaning the single-dish beam size in physical units is too small to be comparable to our work.  Even though the galaxies studied by \citet{gereb16,gereb18} are part of the GASS sample, because they are selected to be gas-rich systems, their lack of difference in single-dish and interferometric \HI~mass measurements is entirely consistent with the results of TNG100 (i.e.~the upper percentiles do not change significantly between the left and right panels of Fig.~\ref{fig:NeutralFrac}).  To settle if the dip is at all reflective of reality with single-dish \HI~data alone would require a survey strategy with similar numbers to GASS for statistics, with longer exposures for a higher detection fraction, and a smaller beam size to reduce any potential of including unbound gas.  This is a challenging task that requires extensive future work.\\

\item The dip occurs at the mass scale where galaxies are understood to transition from being predominantly regulated by stellar feedback to AGN feedback \citep[e.g.][]{croton06}, which is also associated with changes to morphology and star formation activity in observations \citep[e.g.][]{kauffmann03}.  In the original Illustris simulation, overly powerful AGN feedback led to galaxy groups being too gas-poor \citep{genel14}.  A new feedback scheme was introduced in TNG to alleviate this issue \citep{weinberger17}, and while largely successful in its purpose, a similar dip is seen in the gas fraction of TNG haloes of mass $\gtrsim \! 10^{12}\,{\rm M}_{\odot}$, which has explicitly been linked to AGN feedback \citep[see fig.~8 of][]{pillepich18a}.  Note, the impact of AGN feedback on gas fractions as a function of galaxy \emph{stellar mass} was not assessed in those previous works.  Satellites' gas fractions could be more noticeably affected by AGN feedback, as satellites lack the same ability as centrals to re-accrete the ejected gas; instead, what is ejected from a satellite may end up subsequently cooling onto the central galaxy.  Modifying the AGN feedback in TNG might plausibly change the number of galaxies at the knee of the galaxy stellar mass function as well (the same mass scale as the neutral-fraction dip).  While TNG100 recovers the observed knee quite well, there is room within observational uncertainty for this to be altered by $\sim$0.15\,dex \citep[cf.~fig.~14 of][]{pillepich18b}.
\end{enumerate}
Regardless of whether this dip exists in reality or not (which we stress that we can not currently definitively determine), one should not undermine the fact that the simulation results generally align well with observations.  We are, therefore, well positioned with TNG100 to not only investigate the gas properties of galaxies in different phases, but also the effect environment has on them.\\

\begin{figure}
\centering
\includegraphics[width=\textwidth]{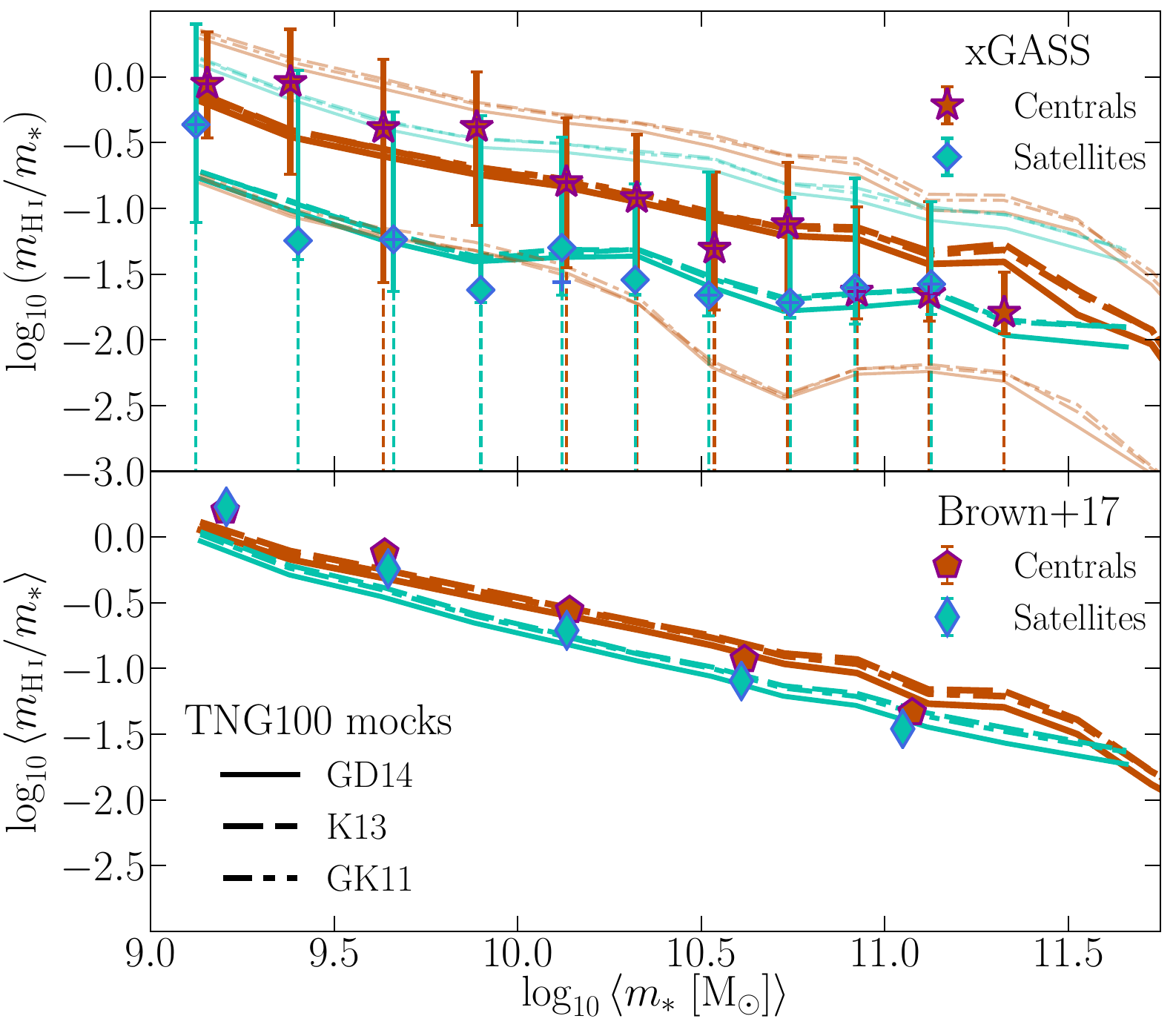}
\caption{Top panel: As for the right panel of Fig.~\ref{fig:NeutralFrac}, but now only considering \emph{atomic} hydrogen.  Line styles are used to differentiate between prescriptions for the partitioning of neutral gas into its atomic and molecular components in TNG100.  The complete xGASS representative sample \citep{catinella18} is used for the observational data.  Bottom panel: The \emph{mean} \HI~fraction of TNG100 galaxies, as measured for best comparison to the stacked ALFALFA data of \citet{brown17} -- see Section \ref{ssec:alfalfa}.  While error bars in the top panel cover percentile ranges, those in the bottom panel are errors \emph{on the mean}, and are generally smaller than the points themselves.}
\label{fig:HIFrac}
\end{figure}

In Fig.~\ref{fig:HIFrac}, we present the \HI~fraction of TNG100 satellites and centrals -- following the three methods for breaking neutral hydrogen into its atomic and molecular components, described in Section \ref{ssec:hih2} and Appendix \ref{app:equations} -- and compare them to both xGASS and ALFALFA.  The difference in results for the three prescriptions is almost entirely negligible at all mass scales \citep[qualitatively in line with][]{diemer18}.  These are also generally in close agreement with observations, especially satellites (TNG satellite lines are typically within $\sim$0.07 and $\sim$0.15\,dex of the xGASS median and ALFALFA mean, respectively, across the full mass range).  Given Fig.~\ref{fig:NeutralFrac} and knowing \HI~accounts for the majority of neutral gas, neither of these findings are particularly surprising.  Again, the only main deviation is that centrals are systematically gas-rich at the high-mass end.  We now see this not only in the comparison to xGASS data (top panel), but also in the stacked ALFALFA result that shows the mean relation (bottom panel).  We note that this contrasts with results from the EAGLE and {\sc Mufasa} simulations, where they instead produced galaxies at the high-mass end that were systematically gas-poor (see fig.~13 of \citealt{saintonge17}; also see \citealt{bahe16,crain17,dave17}).

Comparing galaxies at low masses in the two panels, Fig.~\ref{fig:HIFrac} highlights how important the beam size is in measuring \HI~fractions.  The mean \HI~fraction for satellites with $m_*\!<\!10^{10}\,{\rm M}_{\odot}$ from ALFALFA is \emph{significantly} higher than the median from xGASS (true for both the data and TNG100 mock measurements).  While the log of the mean should almost always be higher than the log median, this is not enough to account for the $>$0.8\,dex difference between the TNG100 mocks in the lowest-mass bin.  Because these galaxies in ALFALFA are observed at $z\!>\!0.02$, whereas those in xGASS are at $z\!<\!0.02$, the physical beam size for the ALFALFA galaxies is larger, meaning more diffuse gas around the galaxies is captured, and the likelihood of confusion is increased.  The lack of beam response imposed in the ALFAFA mock (Section \ref{ssec:alfalfa}) amplifies this effect.  This also contributes to the artificial result that satellites appear to have almost equal \HI~fractions as centrals on average, especially as one goes to lower stellar masses (bottom panel of Fig.~\ref{fig:HIFrac}); the vast majority of centrals at these masses are isolated, while satellites, by definition, must have at least one galaxy nearby, and will therefore typically live in more-massive haloes.  Thus, measurements of satellites' \HI~content are more biased towards including diffuse gas and/or gas of their neighbours, an effect that is stronger with larger beam size.


\section{Physical interpretation of satellites' \HI~content}
\label{sec:SatEnv}

In the previous section, we covered that satellite galaxies have systematically less cold gas compared to centrals of the same stellar mass, both in observations and TNG100.  In this section, we show and discuss what physically drives the \HI~content of satellite galaxies down.

\subsection{Impact of parent halo mass on satellites}
\label{ssec:halomass}

\begin{figure*}
\centering
\includegraphics[width=\textwidth]{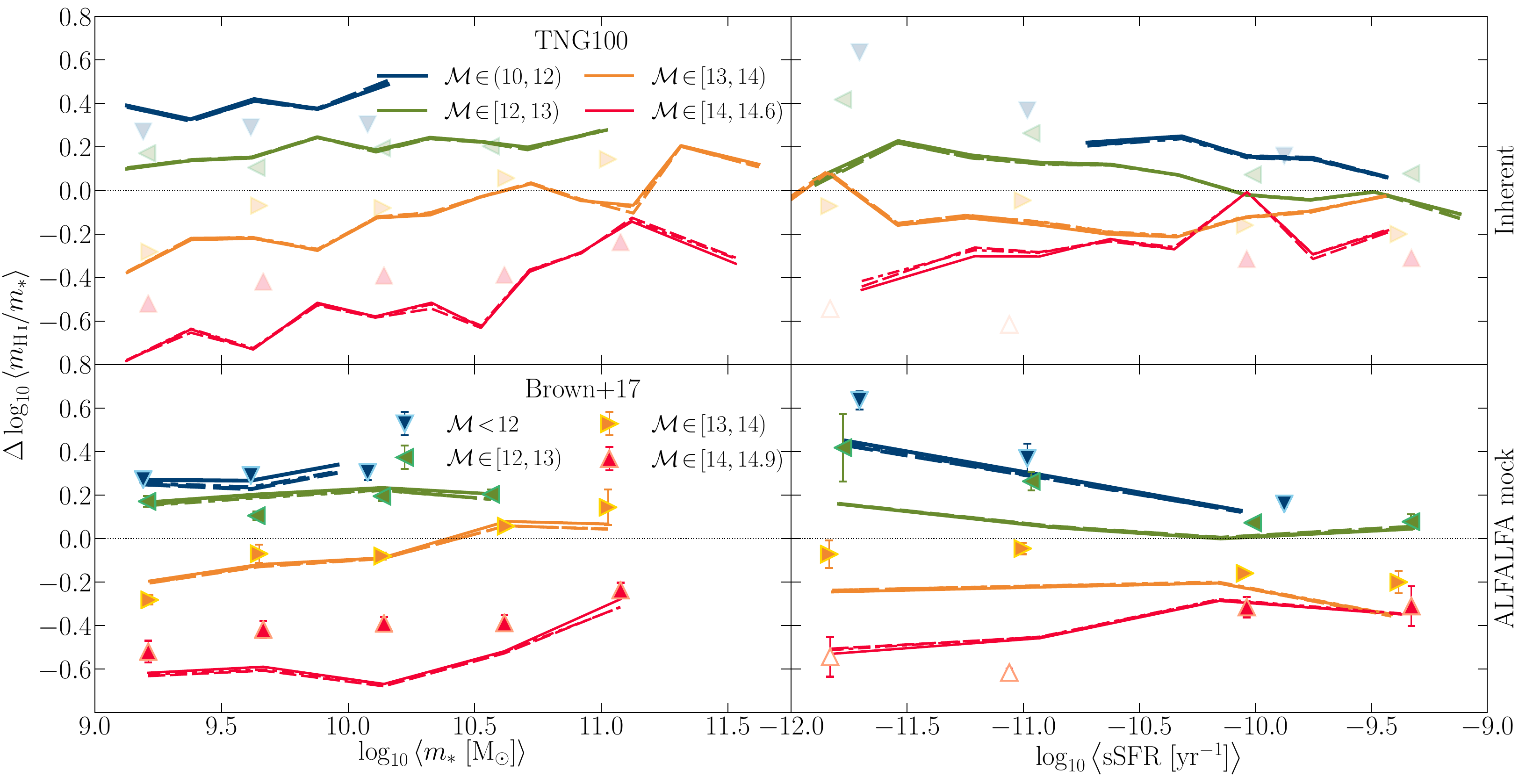}
\caption{\emph{Difference} in the logarithm of the \emph{mean} \HI~fraction of satellite galaxies hosted in certain halo mass bins [as per the legend, where $\mathcal{M}\!\equiv\!\log_{10}\left(M_{\rm 200c,parent}/{\rm M}_{\odot}\right)$] to that of the overall satellite population at the same stellar mass (left panels -- cf.~the lower panel of Fig.~\ref{fig:HIFrac}) and specific star formation rate (right panels).  Line style indicates the \HI/\Htwo~prescription as per previous figures, although no noteworthy difference between these exists in this case.  Inherent TNG100 galaxy properties in the top panels follow the method outlined in Section \ref{ssec:finding}, while the mock-observed properties in the bottom panel follow Section \ref{ssec:alfalfa}.  The latter accounts for the square aperture of ALFALFA, the survey's redshift distribution, and observational uncertainty in stellar mass, parent halo mass, and SFR.  Data from \HI~spectral stacking from the ALFALFA survey \citep{brown17} are shown in both the upper and lower panels to guide the eye, but are only fairly compared to TNG100 results in the lower panels.  Empty points imply a clean detection could not be made for the stack (and thus are upper limits).  Bins of width 0.2\,dex for $m_*$ and 0.3\,dex for sSFR are used for TNG100 galaxies in the top panels.  In the bottom panels, we use bins that best match the observational data (in number, mean stellar mass / sSFR, and width).  Galaxies with $m_*\!>\!10^{11.5}\,{\rm M}_{\odot}$ have been excluded from the bottom panels.  Inherent SFRs are based on instantaneous SFRs of gas cells, while the mock uses the birth mass of stellar particles within a given age to best compare to observations (see Section \ref{ssec:alfalfa}).}
\label{fig:SatEnv}
\end{figure*}

It is well established that the \HI~content of satellite galaxies depends on their environment (Section \ref{sec:intro}).  One means of quantifying environment is with parent halo mass; higher-mass haloes encompass regions of stronger over-density. \citet{brown17} showed how the average \HI~fractions of satellites change with parent halo mass for both fixed stellar mass and fixed specific star formation rate.  Using the same stacked ALFALFA data (Section \ref{ssec:alfalfa}) in conjunction with the {\sc Dark Sage} semi-analytic model of galaxy formation \citep{stevens16}, \citet{sb17} showed explicitly that ram-pressure stripping of the interstellar media of satellites could be held almost entirely responsible for the observed effects (although results from the \citealt{gonzalez14} version of {\sc galform} published in \citealt{brown17} imply this does not necessarily \emph{have} to be the case).  The strength of ram pressure strongly depends on local density \citep{gunn72}, meaning if two satellites of the same mass fall into haloes of different mass, the one in the higher-mass halo will be stripped of its \HI~more efficiently.  Moreover, because lower-density gas in the galaxy's outskirts is more susceptible to stripping, the rates at which \HI~and \Htwo~(or star-forming gas) are stripped are not the same \citep[see][]{fumagalli09}.  In fact, \citet{sb17} found that cold-gas stripping had a negligible effect on SFR, and instead starvation was almost exclusively the cause for their decline relative to centrals \citep[qualitatively similar results from other models have been discussed in][]{cora18,xie18}.  This helps physically explain where the observed effect at fixed sSFR comes from.

Alongside the stacked ALFALFA data of \citet{brown17}, in the bottom panels of Fig.~\ref{fig:SatEnv}, we present the difference in mean \HI~fraction for satellites as a function of fixed stellar mass and sSFR for bins of parent halo mass in TNG100.  The simulation follows the observed trends with parent halo mass and stellar mass remarkably well, with typical deviations between the data points and simulation lines of $\sim$0.06\,dex. This means \emph{the strength of stripping in TNG100 is close to what happens in reality}.  Again, the importance of mock-observing the simulations is clear -- the impact of parent halo mass on satellites'~\HI~fractions at fixed stellar mass is dampened in the mock versus the inherent galaxy properties, whereas the impact is enhanced at fixed sSFR (cf.~the top panels of Fig.~\ref{fig:SatEnv}).  For fixed sSFR, the significant difference in the mock relative to the inherent measurements arises not just because of how \HI~mass is measured, but also how SFR is measured (see Section \ref{ssec:alfalfa}). The apparent (effectively artificial) enhancement of the impact of environment is not enough to bring TNG100 as in line with observations at fixed sSFR as it is with fixed stellar mass, however, with deviations from the data at fixed sSFR of order $\sim$0.07\,dex.  As we will discuss in the next subsection, this indicates that \HI~and star formation are more closely coupled in TNG100 galaxies than they are observed to be.


\subsection{Where and how satellites' \HI~is stripped}
\label{ssec:phase}

Nominally, ram-pressure stripping is controlled by three factors: (i) the density of the ambient medium the satellite is moving through, (ii) the relative velocity of the satellite to that medium, and (iii) the shape and depth of the satellite's gravitational potential well \citep[cf.][]{gunn72}.  The former two factors mean that the amount of stripping a satellite experiences will depend on its orbit in the halo.  After its first pericentric passage, the stripping strength a satellite experiences will decrease.  Only on subsequent passages where the orbit has evolved to have a smaller pericentre should this strength be exceeded again \citep[see, e.g.,][]{bruggen08}.  

A common diagnostic used to understand satellites' gas content in terms of their orbits is a phase-space diagram, which shows the relative position and velocity of satellites within their groups \citep*[e.g.][]{mahajan11,jaffe16,yoon17}.  If the above picture of ram pressure holds, galaxies at low radius and high-magnitude velocities should be biased towards being gas-poor.  Gas-rich and gas-normal systems should be found more often at large radii and low-magnitude velocities.  But gas-poor systems should be found in these parts of phase space too, as no distinction is made (nor can it be generally made, observationally speaking) between the \emph{number} of orbits a satellite has had from this diagnostic alone.

Observationally, it is the line-of-sight velocity and orthogonal displacement that are used for phase-space diagrams. For a hydrodynamic simulation, where we have full 6D information, we can plot the actual radial distance from the halo centre and its derivative (i.e.~its velocity component away from the centre).  This not only eliminates projection effects, but also immediately tells us whether the galaxy is moving towards or away from its pericentre.  This helps to break some of the visual degeneracy between galaxies that are freshly infallen and those that have already completed an orbit(s).  That is, new infallers should exclusively have negative velocities and be at large radii (occupying the bottom-right hand corner of the diagram).

\begin{figure*}
\centering
\includegraphics[width=\textwidth]{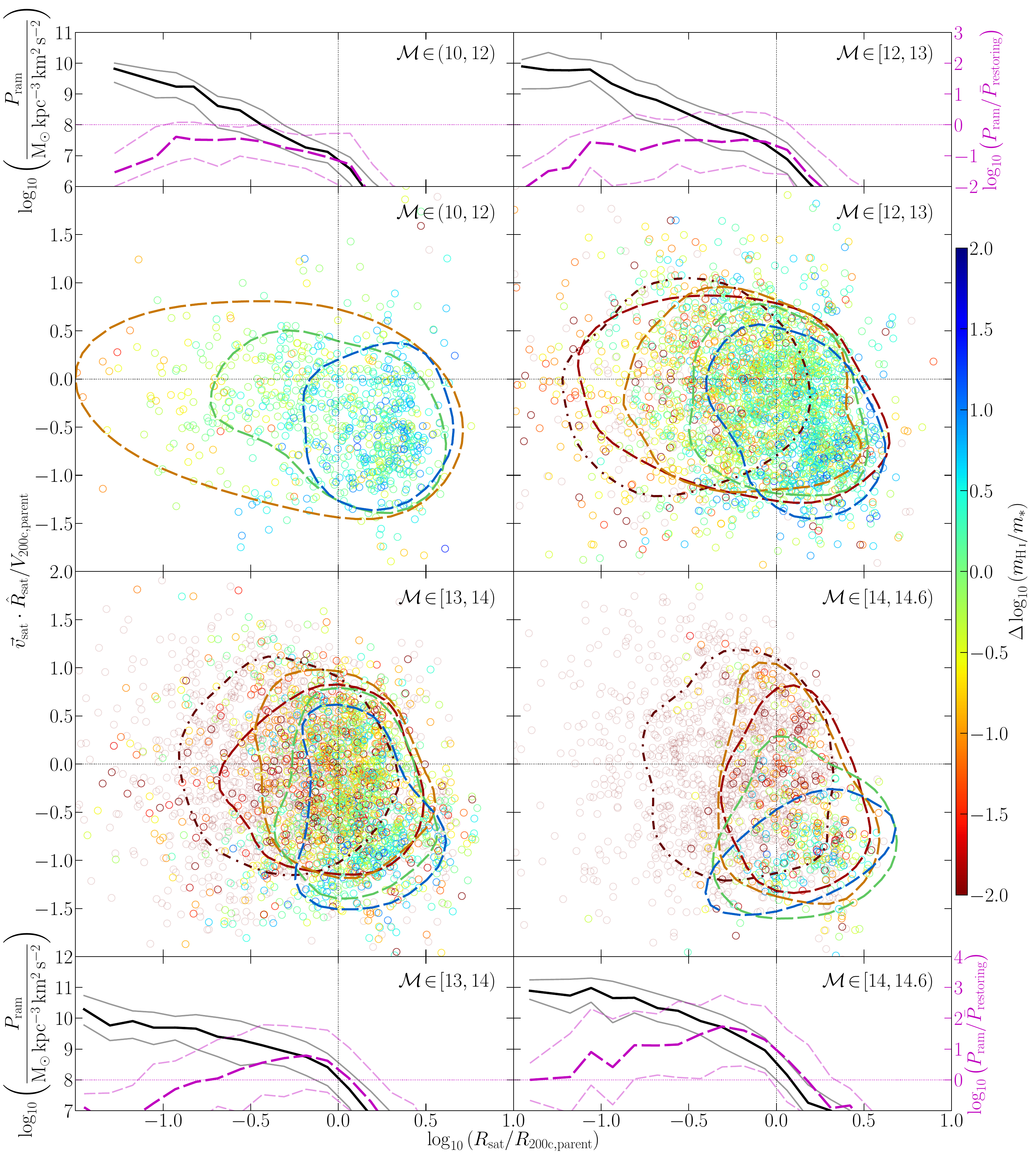}
\caption{Central four panels: phase-space diagrams for all satellites in TNG100 at \zo, i.e.~radial distance from the group central's position against the satellite's relative velocity component along that radial vector.  Each panel corresponds to a halo-mass bin, specified in the corners.  Vertical dotted lines separate galaxies inside and outside the halo virial radius.  Horizontal dashed lines separate infalling (negative velocity) from uprising (positive velocity) galaxies.  The colour of each circle represents the galaxy's \HI~fraction from the \citetalias{gd14} prescription \emph{relative} to the median for galaxies of its stellar mass.  Contours encapsulate 68\% of galaxies in bins of $\Delta \log_{10}\left(m_{\rm H\,{\LARGE{\textsc i}}} / m_* \right)$ (typically of width 1\,dex), each coloured consistently with the bin's centre.  The dot-dashed contours exclusively account for subhaloes that have no gas cells associated with them at all.  Each contour represents a population of $>\!50$ galaxies.  All galaxy/halo properties in this figure are inherent.  Top and bottom panels: strength of ram-pressure stripping for the galaxies in the same halo mass bins.  Solid lines track absolute strength, while dashed lines are normalised by the average gravitational restoring force per unit area of the galaxy disc.  Thick lines are medians, thin lines are 16$^{\rm th}$ and 84$^{\rm th}$ percentiles.}
\label{fig:PhaseSpace}
\end{figure*}

In the central panels of Fig.~\ref{fig:PhaseSpace}, we present \emph{stacked} phase-space diagrams for TNG100 haloes, using the same halo-mass bins as in Section \ref{ssec:halomass}.  That is, all satellites belonging to haloes in each mass bin are plotted, where the position and velocity of each satellite is normalised by the virial radius and velocity of its host halo, respectively.  The relative \HI~richness of each galaxy (measured as the difference in \HI~fraction from the median of all galaxies in the simulation at the same stellar mass) is indicated by colour.  We use contours to help clarify where galaxies of a fixed \HI-richness lie in phase space.

The results shown in Fig.~\ref{fig:PhaseSpace} match the expectations of ram-pressure stripping described above.  In the highest halo-mass bin ($M_{\rm 200c} \! \geq \! 10^{14}\,{\rm M}_{\odot}$), nearly all galaxies close to the halo centre are stripped \emph{completely} of gas (not just \HI).  Only galaxies near or beyond the virial radius have significant \HI~masses.  Of those with positive velocities (i.e.~galaxies that have completed at least one orbit), the majority are gas-poor.  By selecting galaxies that are gas-rich, we find they can exist in these clusters at similar radii, but they almost exclusively have negative velocities, implying they must only have very recently become satellites, and therefore have only had minimal exposure to stripping effects thus far.  Stepping down in halo mass, we find that galaxies of all \HI~fractions start to occupy more and more of phase space, implying satellites can retain some of their gas even after multiple orbits.  Gas-rich galaxies are still typically found at larger radii and more-negative velocities though.

In the upper and lower panels of Fig.~\ref{fig:PhaseSpace}, we explicitly show the range of strength of ram pressure the galaxies are experiencing as a function of their halo-centric distance.  To calculate the strength of ram pressure on each galaxy, we need to know the local density of the gas medium it is moving through, and the relative speed at which it is moving, i.e.
\begin{equation}
P_{\rm ram} = \rho_{\rm IHM}(R_{\rm sat})\, v_{\rm sat}^2
\end{equation}
(IHM = intrahalo medium).  To obtain $\rho_{\rm IHM}(R_{\rm sat})$, we build one-dimensional density profiles for the non-star-forming gas of every central (i.e.~throughout the halo, excluding the mass in the satellites) and interpolate the positions of the satellites.  We take the velocity magnitude of the satellite relative to the central as $v_{\rm sat}$.  We show not only the absolute ram-pressure strength in Fig.~\ref{fig:PhaseSpace}, but also its value normalised by the average gravitational restoring force per unit area of the galaxy, calculated as
\begin{subequations}
\begin{equation}
\bar{P}_{\rm restoring} \equiv 2\, \pi\, G\, \bar{\Sigma}_* \left( \bar{\Sigma}_* + \bar{\Sigma}_{\rm gas} \right)\,,
\end{equation}
\begin{equation}
\bar{\Sigma}_x \equiv \frac{m_{x, \rm disc} }{\pi\, r_{\rm disc}^2}\,,
\end{equation}
\end{subequations}
(see Section \ref{ssec:disc} for disc property definitions).  Our calculations of both $P_{\rm ram}$ and $\bar{P}_{\rm restoring}$ are approximate \citep[cf.~the differences in the calculations for TNG galaxies by][]{yun18}, but they help to illustrate the importance of ram pressure; if $P_{\rm ram} \! \gtrsim \! \bar{P}_{\rm restoring}$, a galaxy is susceptible to having at least some of its gas stripped by ram pressure \citep[cf.][]{gunn72}.  

As expected, the absolute strength of ram pressure increases at lower halo-centric radii and for galaxies in haloes of higher mass in TNG100 (Fig.~\ref{fig:PhaseSpace}).  
For haloes with $M_{\rm 200c}\!<\!10^{12}\,{\rm M}_{\odot}$, $P_{\rm ram}$ only exceeds $\bar{P}_{\rm restoring}$ for a small minority of satellites inside $R_{\rm 200c}$ ($<$16\% -- the 84$^{\rm th}$ percentile of $P_{\rm ram} / \bar{P}_{\rm restoring}$ is less than unity at all radii).  On average, these haloes host 2.14 group members with $m_* \! \geq \! 10^9\,{\rm M}_{\odot}$, meaning most of these satellites are the smaller galaxy of a well-resolved pair.
In the next halo mass bin, even at $R_{\rm 200c}$, a more notable fraction of satellites are susceptible to ram pressure.  Observations of Milky Way ($M_{\rm 200c} \!\simeq\! 10^{12}\,{\rm M}_{\odot}$ -- see \citealt{elahi18} and references therein) satellites are consistent with this picture, in that inside the virial radius, satellites have notably lower \HI~masses than those beyond, for which ram pressure is capable of explaining \citep[see][]{grcevich09,gatto13,spekkens14}.  For haloes with $M_{\rm 200c}\!\geq\!10^{13}\,{\rm M}_{\odot}$ (i.e.~groups and clusters), $P_{\rm ram} \! > \! \bar{P}_{\rm restoring}$ even beyond $R_{\rm 200c}$ for many satellites in TNG100.  The downturn seen in the  $P_{\rm ram}/ \bar{P}_{\rm restoring}$ profiles towards lower radii (despite the continuous rise of $P_{\rm ram}$ in absolute units) is analogous to survival bias.  That is, galaxies that have been stripped will only have small, dense discs remaining.  Effectively by definition, whatever is left must be impervious to ram pressure.  Many galaxies will have already had at least one pericentric passage, meaning the ram pressure they feel now is not the strongest they have felt in their history.


It is important to emphasise that we have not \emph{explicitly} shown that ram-pressure is \emph{definitively} the process that leads to satellites' gas loss in TNG100.  Here, we have simply demonstrated that it is a strong and valid candidate.  This is consistent with the results of \citet{yun18}, who show that a non-negligible fraction of TNG100 satellites in massive groups and clusters exhibit visible signs of ram-pressure stripping in the form of elongated gaseous tails and gas asymmetries.  

As discussed in Section \ref{sec:intro}, other environmental effects like gravitational tides can strip a galaxy of its gas too.  To properly disentangle which phenomena are responsible for the stripping of gas in hydrodynamic-simulation galaxies is non-trivial, as their implementation is entirely implicit.  To thoroughly do this arguably warrants a study in and of itself.  While we leave this detailed analysis of TNG for future work, we refer the reader to \citet{marasco16} for an extended discussion on the processes responsible for satellites' gas loss in the EAGLE simulations, where qualitatively similar trends are found with environment, and the significance of ram pressure is highlighted.


\subsection{When \HI~is stripped versus quenching}
\label{ssec:time}

\begin{figure}
\centering
\includegraphics[width=\textwidth]{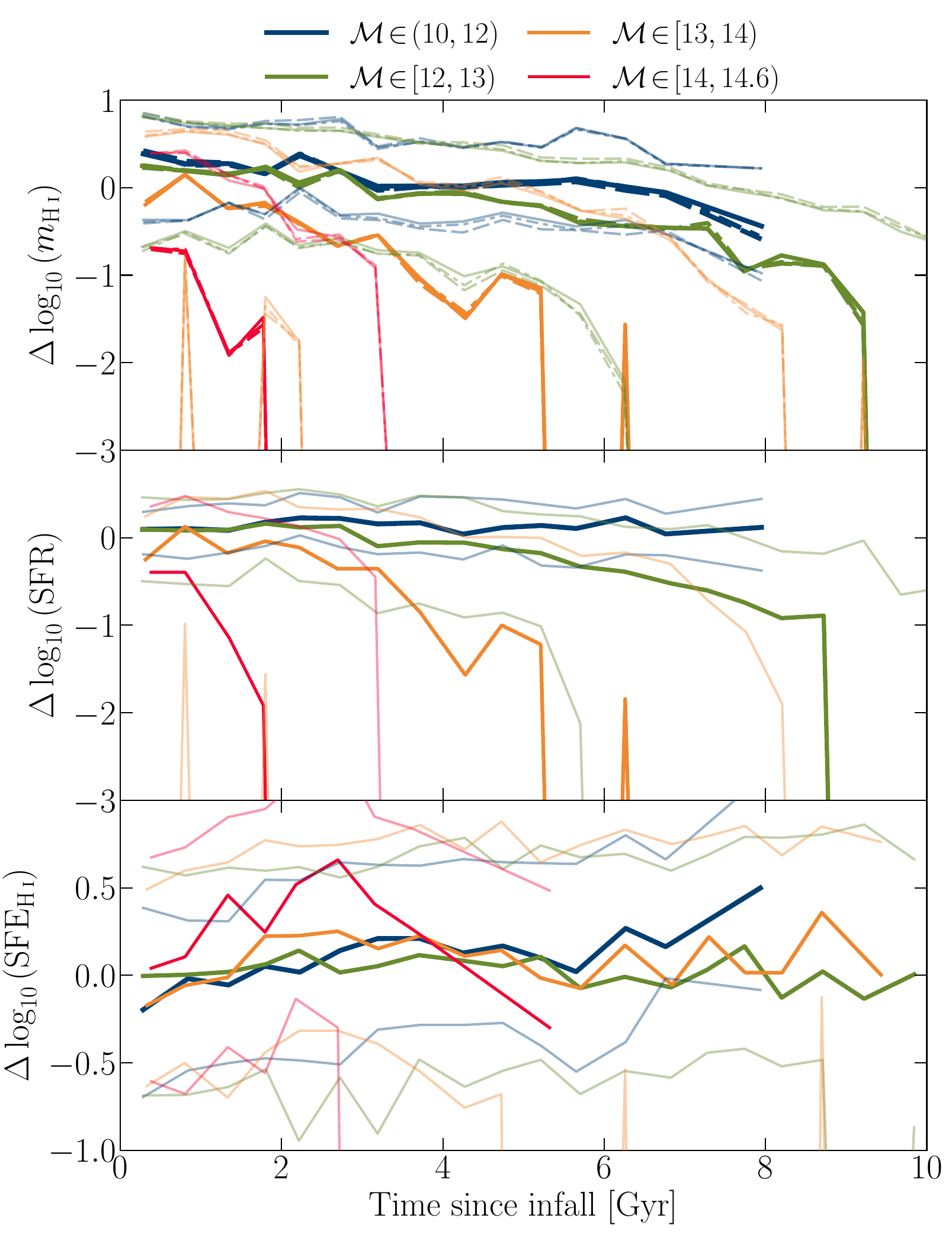}
\caption{Top panel: \HI~mass of TNG100 satellite galaxies at \zo, normalised by the median \HI~mass for all \zo~galaxies at the same stellar mass (the $y$-axis corresponds to the colour bar in Fig.~\ref{fig:PhaseSpace}, all properties are inherent), as a function of time since infall (i.e.~since the snapshot the galaxy was first identified as a satellite by {\sc subfind}).  Satellites are broken into bins of parent halo mass, indicated by colour.  Line style for the top panel corresponds to \HI/\Htwo~prescription (legend in Fig.~\ref{fig:HIFrac}, no differences to note in this figure).  Thick lines are medians, whereas thin, transparent lines are 16$^{\rm th}$ and 84$^{\rm th}$ percentiles; these were built from bins of minimum width 0.25\,Gyr with at least 20 galaxies per bin.  Middle and bottom panels: star formation rates and \HI~star formation efficiencies of TNG100 satellites, respectively, normalised by the median relevant quantity for satellites of their stellar mass in the simulation at \zo.  Contrary to expectation, satellites' \HI~content and SFRs are equally affected by their halo environment in TNG100.}
\label{fig:tinfall}
\end{figure}

It should be generally true that satellites that are more stripped have been in the halo for longer, as their exposure to stripping effects would be maximised (modulo their orbital parameters).  The top panel of Fig.~\ref{fig:tinfall} shows exactly this.  Not only is there a clear decline in the relative \HI~content of satellites with time since infall (defined as the time since the galaxy first became a satellite, i.e.~when {\sc subfind} identified it as not being the main subhalo of a halo), but stripping time-scales are clearly seen to be much shorter for satellites living in higher mass haloes (note that the average time between snapshots is $\sim$140\,Myr).  What we also found is that SFRs of satellites decline with time since infall in an almost identical fashion as \HI~mass -- this is shown in the middle panel of Fig.~\ref{fig:tinfall}.  \emph{This result is contrary to the canonical picture of ram-pressure stripping in which extended \HI~should be lost prior to the quenching of star formation}.  To check this finding, we explicitly calculate the ratio of the satellites' star formation rates to their \HI~mass, sometimes referred to as an `\HI~star formation efficiency', ${\rm SFE}_{\rm H\,{\LARGE{\textsc i}}} \equiv {\rm SFR} / m_{\rm H\,{\LARGE{\textsc i}}}$, and compare how this changes with parent halo mass as a function of time since infall in the bottom panel of Fig.~\ref{fig:tinfall}.  If \HI~were stripped before satellites started quenching, then denser environments with stronger stripping should host satellites with higher ${\rm SFE}_{\rm H\,{\LARGE{\textsc i}}}$.  Instead, there is nearly no environmental dependence seen; only in the highest halo mass bin is there a suggestion of galaxies that have been satellites for $<\!4\,{\rm Gyr}$ having a sightly higher ${\rm SFE}_{\rm H\,{\LARGE{\textsc i}}}$ on average, but this is marginal at best.

\begin{figure}
\centering
\includegraphics[width=\textwidth]{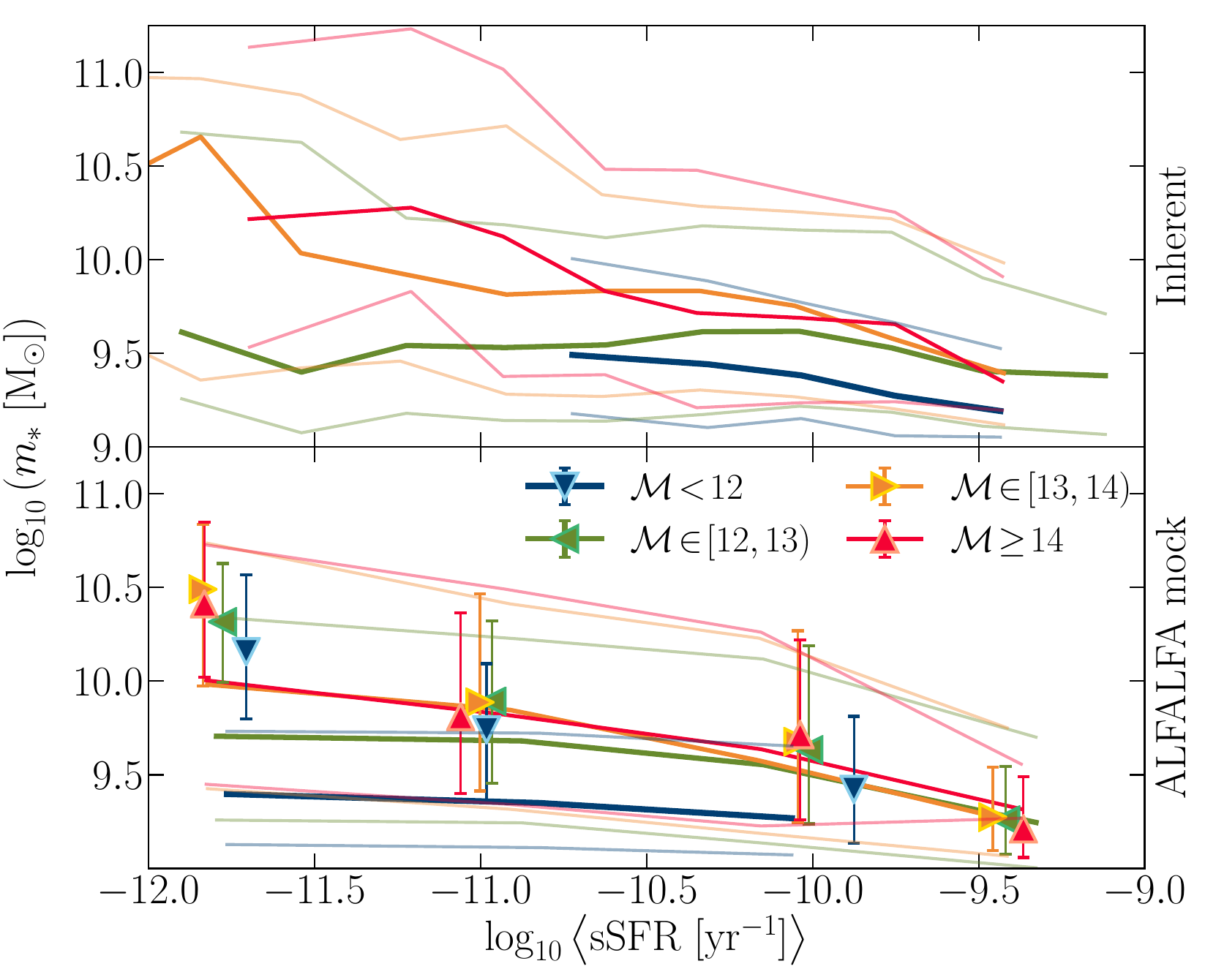}
\caption{Distribution of satellite stellar masses as a function of specific star formation rate for different bins of parent halo mass.  Thick lines are running medians for TNG100.  Thin lines are the 16$^{\rm th}$ and 84$^{\rm th}$ percentiles. Points compared in the bottom panel are data from the \citet{brown17} sample.  The binning strategy matches the right panels of Fig.~\ref{fig:SatEnv}.  Differences between the data and mock in the bottom panel help to reconcile the bottom right panel of Fig.~\ref{fig:SatEnv} with Fig.~\ref{fig:tinfall} (see text in Section \ref{ssec:time} for details).}
\label{fig:ssfr_mstar}
\end{figure}

Of central importance to the result of Fig.~\ref{fig:tinfall} is the gas density threshold for star formation that TNG imposes.  Typical for modern cosmological, hydrodynamic simulations \citep[cf.][]{dubois14,schaye15}, star formation will only take place in TNG gas cells whose densities exceed $\sim$0.1 proton masses per cm$^3$.  In the real Universe, stars form at much higher gas densities.  To raise this threshold would inevitably reduce how extended star formation is in TNG galaxies, and would also make star formation take place more exclusively in molecular-dominant cells.  In principle, this should encourage preferential stripping of \HI, thereby improving agreement with observations in that regard.  This is clearly an area of investigation for future work.

\citet{brown17} used the canonical picture of \HI~preferential stripping to explain the effect of parent halo mass on \HI~fraction at fixed sSFR (Fig.~\ref{fig:SatEnv} of this paper).  As noted above, something else must be driving the difference in TNG100.  We suggest this could be due to a stellar-mass bias in the simulation.  That is, at fixed sSFR, the galaxies in TNG100 and ALFALFA might not have the same typical stellar mass, which would imply they are not necessarily directly comparable galaxy populations.  In Fig.~\ref{fig:ssfr_mstar} we show the running distribution of satellites' stellar masses using the same bins of sSFR and parent halo mass as in Fig.~\ref{fig:SatEnv}.  In TNG100, we find that for halo masses below $10^{14}\,{\rm M}_{\odot}$, the stellar masses of satellites systematically increase at fixed sSFR for increasing halo mass (cf.~the median lines in Fig.~\ref{fig:ssfr_mstar} for ${\rm sSFR} \! \lesssim \! 10^{-10}\,{\rm yr}^{-1}$).  This is true for both the inherent and mock-observed galaxy properties.  Being biased towards higher stellar masses means being biased towards lower \HI~fractions (as seen in Fig.~\ref{fig:HIFrac}).

However, the observational data do \emph{not} appear to be subject to the same bias (or at least certainly not to the same extent as TNG100), as seen by the relative closeness of the points in the bottom panel of Fig.~\ref{fig:ssfr_mstar}.  Our results therefore do not really argue against the canonical stripping picture, but more likely highlight the challenging nature of resolving multi-phase gas stripping in cosmological, hydrodynamic simulations, especially when only done in post-processing; for the phases to be \emph{truly} dynamically separable would require incorporating the phase breakdown into the hydrodynamics scheme of the simulation.  As we will detail in a follow-up paper (Stevens et al.~in prep.), the change in \Htwo~fraction with environment for TNG100 galaxies is very similar to that for \HI.  While observational data are fewer, some are in tension with this; recent observational evidence suggests the molecular content of galaxies' interstellar media is not strongly affected by ram-pressure stripping \citep[but the gas that is stripped can be subject to a phase change -- see][]{lee17,moretti18}.  However, \citet{boselli14c} have demonstrated that galaxies in clusters tend to be \Htwo-poor relative to those in the field of the same stellar mass.  Further investigation is clearly warranted to better grasp how \HI~and \Htwo~in galaxies are relatively affected by ram-pressure stripping.

One aspect of our results we have not investigated to its fullest potential is the effect of the halo finder used to assign central/satellite status and halo masses to galaxies in the observational surveys.  While we made efforts to account for how the \citet{yang07} catalogue makes these decisions in building our mocks from TNG100 (Section \ref{sec:obs}), this does \emph{not} mean we recover the same contamination rates that are present in the survey catalogues; in our mocks, the central and satellite assignments are $>\!98$\% pure in each case (compared to the {\sc subfind} assignments), but in the survey catalogues, satellites only have a purity of $\sim$60\% \citep{campbell15}.


\subsection{Connection to angular momentum}
\label{ssec:qf}

There is a wealth of recent evidence that the \HI~content of galaxies is correlated with their specific angular momenta \citep[see][]{ob15,ob16,lagos17,stevens18,wang18,zoldan18}.  \citet{ob16} frame this connection in terms of the global stability of a disc.  Discs that are larger and/or rotate faster are less prone to gravitational instabilities.  Specific angular momentum captures both these contributions.  Gas in a stable disc is less likely to collapse into molecular clouds and form stars. Thus a greater fraction of baryons in a stable disc should be in an atomic, gaseous state.  Being itself a fundamental property of physics that adheres to conservation laws, specific angular momentum is therefore a natural property to model the \HI~fraction of galaxies in terms of.

\citet{ob16} use a `global disc stability parameter', $q$, that is directly proportional to disc specific angular momentum, $j_{\rm disc}$, and inversely proportional to its mass, to relate to the atomic mass fraction of discs, $f_{\rm atm}$:
\begin{subequations}
\label{eq:qfatm}
\begin{equation}
f_{\rm atm} \equiv \frac{m_{\rm H\,{\LARGE{\textsc i}}, disc}}{X\, m_{\rm disc}} \simeq {\rm min}\left[1,~2.5\,q^{1.12}\right]\,,
\end{equation}
\begin{equation}
q \equiv \frac{\sigma_{\rm H\,{\LARGE{\textsc i}}}\, j_{\rm disc}}{G\, m_{\rm disc}}\,,
\end{equation}
\end{subequations}
where $\sigma_{\rm H\,{\LARGE{\textsc i}}}$ is the radial velocity dispersion of atomic gas in the disc.  The scaling of $f_{\rm atm}$ with $q$ should be much tighter than other galaxy properties, e.g.~stellar mass or SFR \citep[e.g.][]{stevens18}.  A true test of whether a process (e.g. ram-pressure stripping) makes galaxies \HI-poor is therefore to assess its impact on the $q$--$f_{\rm atm}$ relation.

\begin{figure}
\centering
\includegraphics[width=\textwidth]{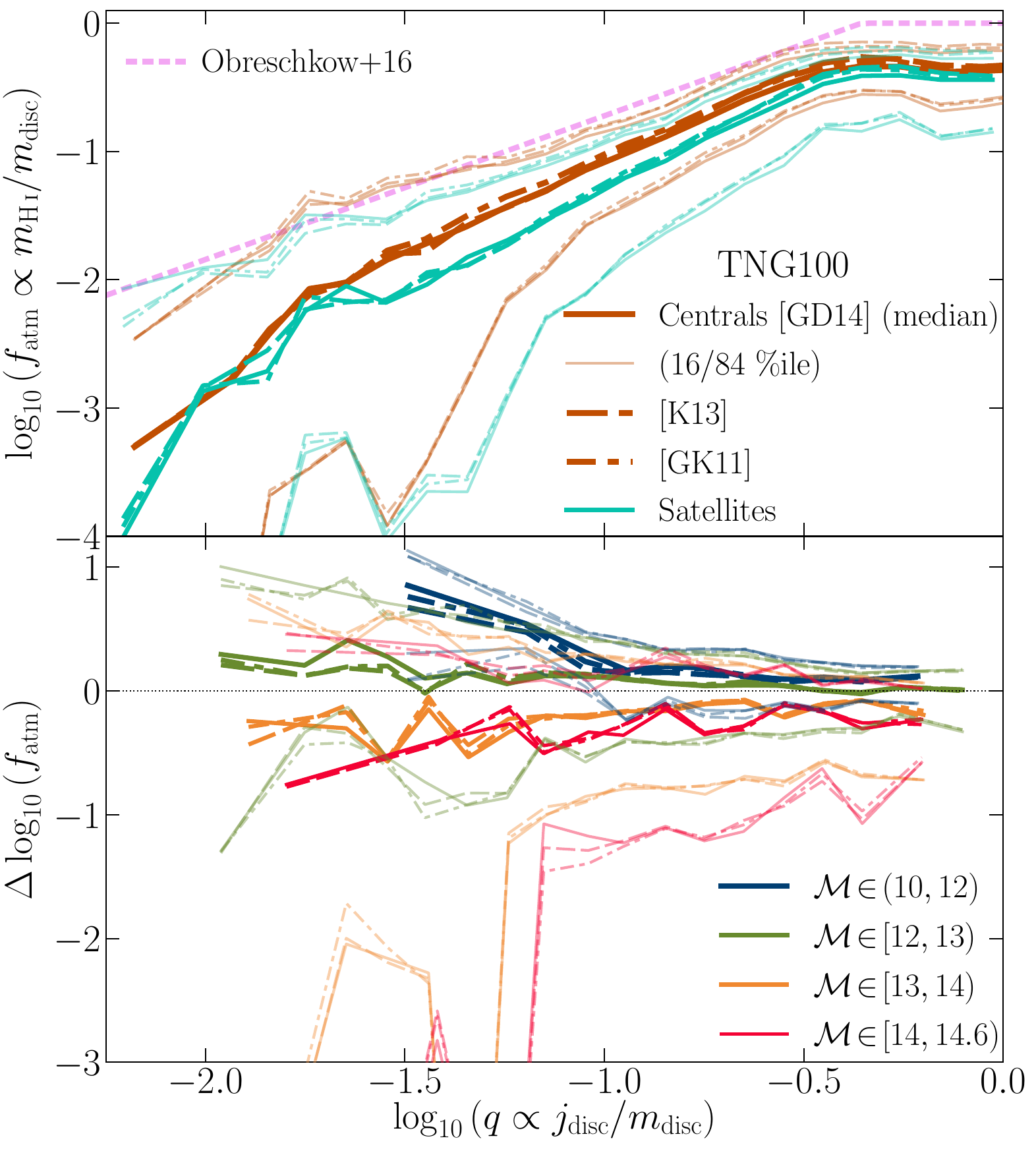}
\caption{Atomic disc fraction in terms of the `global disc stability' parameter for TNG100 galaxies at \zo~that have $m_{\rm H\,{\LARGE{\textsc i}} + H_2,\,disc} \! > \! 0$.  Compared is the analytic prediction of \citet{ob16}.  The bottom panel compares the \emph{difference} in atomic-gas fraction for satellites in denoted parent halo mass bins to the median for all satellites at the same $q$.  Running medians and percentiles are calculated in bins of minimum width 0.1\,dex in $\log_{10}(q)$, each with a minimum of 20 galaxies.  See Section \ref{ssec:qf}.}
\label{fig:qf}
\end{figure}

The top panel of Fig.~\ref{fig:qf} shows the $q$--$f_{\rm atm}$ relation for TNG100 centrals and satellites.  Only galaxies with non-zero neutral gas disc masses are included; were we to not exclude zero-gas galaxies, Fig.~\ref{fig:qf} would not be particularly informative, as the majority of satellites with $q\!<\!10^{-1}$ would have $f_{\rm atm} \! = \! 0$ (cf.~the results of \citealt{stevens18}).  $\sigma_{\rm H\,{\LARGE{\textsc i}}}$ is calculated by isolating rotationally supported gas cells \citep[using the criteria of][]{mitchell18} that are also predominantly atomic, gridding the cells in two dimensions after rotating the galaxy to be face-on, calculating the standard deviation of radial velocities in each grid cell, summing this in quadrature with the thermal velocity dispersion of that gas, then taking a mass-weighted mean of those values.  Neutral gas and stars contribute to the total mass and specific angular momentum of the disc (see Section \ref{ssec:disc}).  This plot highlights the broad applicability of the \citet{ob16} analytic model -- not only does it run approximately parallel to TNG100, but it has also been shown to hold in observations and semi-analytic models of galaxy formation \citep{ob16,lutz18,stevens18}.  

We note that the median $\sigma_{\rm H\,{\LARGE{\textsc i}}}$ for TNG100 galaxies is $\sim$16\,km\,s$^{-1}$, with 68\% of galaxies in the range $(11,28.5)$\,km\,s$^{-1}$.  These numbers only account for galaxies with enough gas cells in the disc to meaningfully measure their dispersion.  For galaxies that had insufficient gas to measure $\sigma_{\rm H\,{\LARGE{\textsc i}}}$ (11.5\% of included cases), we approximate $\sigma_{\rm H\,{\LARGE{\textsc i}}}$ by setting it to the median value for all galaxies that did have it successfully measured (i.e.~16 km\,s$^{-1}$). 
In testing, we found if we manually set all $\sigma_{\rm H\,{\LARGE{\textsc i}}}$ to a constant 10\,km\,s$^{-1}$ \citep[{\' a} la][]{ob16}, then the median lines in Fig.~\ref{fig:qf} align with the analytic model.  Systematic differences from \citet{ob16} in Fig.~\ref{fig:qf} are therefore more down to systematic differences between $\sigma_{\rm H\,{\LARGE{\textsc i}}}$ of TNG100 galaxies and observations \citep[cf.][]{tamburro09}, and less a signal that the angular momentum--gas fraction connection is different to the analytic model.

For TNG100, the distinction between satellites and centrals when it comes to $f_{\rm atm}$ at fixed $q \! \in \! (10^{-2}, 10^{-1})$ is $\sim$0.25\,dex (cf.~the median lines in the top panel of Fig.~\ref{fig:qf}). This is in contrast to \citet{stevens18}, who find a starker separation between centrals and satellites in the {\sc Dark Sage} semi-analytic model, when $q\!\lesssim\!10^{-1}$ and $q\!\gtrsim\!10^{-0.6}$ (see their fig.~7).  In line with this, deviations in $f_{\rm atm}$ at high $q$ for satellites hosted in haloes of different mass (bottom panel of Fig.~\ref{fig:qf}) are \emph{much} smaller in TNG100 -- for comparison, \citet{stevens18} found variations in $f_{\rm atm}$ well in excess of 2\,dex at $q \! \gtrsim \! 10^{-0.6}$.  Galaxies at high $q$ tend to be lower in mass \citep[$q \! \approxprop \! m^{-1/3}$ -- see section 5.2 of][]{stevens18} and therefore should be more susceptible to stripping.  The stripping of \HI~from high-$q$ TNG satellites must be accompanied by a non-negligible loss in specific angular momentum to explain the relatively weak environmental dependence on $f_{\rm atm}$.  At low $q$, results with TNG100 return to qualitative consistency with \citet{stevens18}, in that the effect of parent halo mass is clearly visible on satellites (TNG100 probes even lower values of $q$, where the effect becomes progressively stronger).  These are typically large galaxies, with extended \HI~that is susceptible to stripping, but with a central bulge whose gravitational influence helps to keep gas that is deeper in the galaxy invulnerable.

A key reason why a hydrodynamic simulation and a semi-analytic model would differ in their predictions for the effect of environment on the $q$--$f_{\rm atm}$ relation comes with the need for semi-analytic models to describe \emph{all} environmental effects on baryons explicitly.  For TNG, these instead occur naturally through solving the equations of gravity and hydrodynamics for individual baryonic cells/particles.  Tidal interactions, for example, can strip both gas \emph{and} stars, and can also torque galaxies.  While intrinsically accounted for here, this is often not modelled in semi-analytics \citep[and certainly was not in][]{stevens18}.  Galaxies that are deep in a halo's potential are likely to experience not only stronger ram pressure, but also stronger tidal interactions.  Satellites of high $q$ in TNG100 can still be experiencing strong ram pressure, but the addition of tidal interactions and galaxy harassment could conspire to keep $q$ and $f_{\rm atm}$ tightly related.  This could be tested by adding a prescription for tidal stripping to the semi-analytic model and comparing its output with the prescription turned on and off.


\section{Context and summary of results}
\label{sec:conc}

Using the TNG100 run from `The Next Generation' Illustris suite of cosmological, magnetohydrodynamic simulations \citep{nelson18,pillepich18b,marinacci18,naiman18,springel18}, we have investigated the role that environment (namely central/satellite status and host halo mass) plays in shaping the atomic-hydrogen content of galaxies.  Building from the works of \citet{lagos15b} and \citet{diemer18}, we have developed a comprehensive method for post-processing the gas cells in the simulation to recover their multiphase (i.e.~atomic and molecular) components. We have not only studied the inherent \HI~properties of galaxies (i.e.~integrating the mass of bound gas in galaxies) but also mock-observed them in \HI~to facilitate an accurate comparison with modern surveys at $z \! \simeq \! 0$, including data from the xGASS survey \citep{catinella18} and ALFALFA \citep[specifically, spectrally stacked data from][]{brown17}.  In practice, for the mocks, we have explicitly reproduced the galaxy redshift distributions as a function of stellar mass, adopted consistent apertures or accounted for beam response for the measurements of the gas quantities, included errors in the stellar-mass and SFR measurements, accounted for the time-scale that observed SFRs probe, and reassigned host halo masses based on abundance matching. Our main results are summarised as follows.

The neutral gas fraction of galaxies as a function of stellar mass tracks results from xGASS-CO remarkably closely (order tens of per cent different), for both satellites and centrals independently (right panel of Fig.~\ref{fig:NeutralFrac}).  This is in spite of the fact that the inherent neutral fractions of TNG100 galaxies show a dip around the knee of the stellar mass function, which is likely associated with the onset of AGN feedback (left panel of Fig.~\ref{fig:NeutralFrac}).  The dip is stronger for satellites, as they are less able to reattain gas lost through feedback than centrals, due to their relative lack of gravitational influence.  This feature is washed away when the galaxies are mock-observed; this results from the large beam size imposed to match that of Arecibo (3.5-arcmin FWHM, observing galaxies at $z\!<\!0.05$), which can include the mass of gas cells that are not bound to the galaxies but are along the same lines of sight with similar line-of-sight velocities.  The ongoing {\sc Wallaby} survey should help to shed light on this product of TNG100, as it will measure the \HI~content of hundreds of thousands of galaxies, with several times finer resolution than Arecibo \citep[see][]{duffy12}.

In applying three prescriptions for the \HI/\Htwo~breakdown of gas in the simulation \citepalias{gk11,k13,gd14}, we find little difference in the trends of \HI~fraction with stellar mass in TNG100.  In all cases, these align with the running median and 84$^{\rm th}$ percentile from xGASS (comparison of the 16$^{\rm th}$ percentile is limited by non-detections in the data), and the mean trend found from stacking the 21-cm spectra from ALFALFA galaxies (Fig.~\ref{fig:HIFrac}).  We further find the mean variation in satellite's \HI~fractions with parent halo mass in TNG100, as a function of both stellar mass and specific star formation rate, closely tracks the ALFALFA data (again, within tens of per cent -- Fig.~\ref{fig:SatEnv}).  

Similar analysis comparing the same data was conducted by \citet{sb17} using the {\sc Dark Sage} semi-analytic model \citep{stevens16}.  The most significant point of difference with a semi-analytic model is that all environmental processes must be explicitly prescribed.  While that model produced satellites at \zo~whose \HI~was \emph{relatively} affected by ram-pressure stripping similarly to what is observed, the \emph{absolute} strength of stripping appeared to be too strong; i.e.~the satellites appeared systematically gas-poor, despite the centrals being calibrated to match the observational data.  However, contributions from neither diffuse halo \HI~nor `confusion' galaxies in the same line of sight were considered in that work.  In this paper, we have shown that mock-observing simulated galaxies is paramount to the comparison with survey data.  Significantly more \HI~than is gravitationally bound to low-mass galaxies is included in their \HI~mass measurement.  Our results therefore offer a potential solution for the apparent difficulty semi-analytic models had with recovering both the absolute and relative effects of ram-pressure stripping simultaneously \citep[also cf.][]{brown17,cora18}.

What is especially impressive about the TNG results is that no information about the cold-gas content of galaxies was included in the simulation's calibration \citep[see][]{pillepich18a}.  All results presented in this paper are therefore either predictions or \emph{post}dictions (in the instance the data came first).  This is equally true for results from the EAGLE simulations \citep[see][]{crain15,crain17,schaye15,bahe16,marasco16}.

In using phase-space diagrams, we have shown how \HI-rich galaxies in haloes lie exclusively at large radii and negative radial velocities, implying they have only recently become satellites (Fig.~\ref{fig:PhaseSpace}).  At lower halo masses, gas-normal galaxies can extend all the way into the halo centre, and gas-rich systems are more likely to retain their gas for more than one orbit.  In higher-mass haloes, a greater fraction of satellites are devoid of \HI, with their occupancy extending to increasing radii.  All of this falls in line with expectations and previous studies of ram-pressure stripping \citep[see][]{gunn72,yoon17}.

We have found the \HI~deficit of a satellite galaxy (compared to the median for all galaxies of the same stellar mass) has the same dependence on its time since infall as its star formation rate does (Fig.~\ref{fig:tinfall}).  This contrasts with the narrative based on observational data that \HI~is preferentially removed by ram-pressure stripping due to its relative extension throughout a galaxy, while star-forming (or molecular) gas should be relatively impervious, as it resides deeper in the galaxy's potential well.  Part of the reason why TNG100 still recovers an environmental dependence for \HI~fraction at fixed sSFR is from a bias at fixed sSFR of lower-mass haloes hosting lower-stellar-mass satellites, which is not present in the observational sample.  This highlights the limitations of even the most advanced cosmological simulations in recovering how the multiple phases of the ISM are affected by their environment; to do this thoroughly will require folding the \HI/\Htwo~breakdown into the simulation itself, and accounting for this in the hydrodynamics scheme.  This would not only require the additional modelling of many complex physical processes (dust, radiative transfer), but it would be yet another computationally taxing aspect of the simulation.

Finally, we have investigated how ram-pressure stripping affects the connection between the \HI~fraction and angular momentum of a galaxy disc (Fig.~\ref{fig:qf}).  We find that both centrals and satellites fall equally in line with theoretical expectation \citep{ob16}, modulo a systematic difference that we associate with a discrepancy in the velocity dispersion of \HI~in TNG100 discs compared to observations.  Only at low values of $j_{\rm disc} / m_{\rm disc}$ do we find a strong environmental dependence on \HI~fraction.  This contrasts with semi-analytic predictions, where \citet{stevens18} suggested that stripping should impact \HI~fractions the most at high values of $j_{\rm disc} / m_{\rm disc}$.  The difference in predictions could be due to the semi-analytic model not treating the stripping of stars at all, and could also be affected by our hard cut of $m_* \! \geq \! 10^9\,{\rm M}_{\odot}$ (\citealt{stevens18} included galaxies below this mass).

In future work (Stevens et al. in prep.), we will study the \Htwo~properties of TNG galaxies, comparing closely to xCOLD GASS \citep{saintonge17}.


\section*{Acknowledgements}
All plots in this paper were built with the {\sc matplotlib} package for {\sc python} \citep{hunter07}.  
ARHS thanks G.~Kauffmann for funding towards visits to MPA that helped facilitate some of this work, plus L.~Cortese and D.~Obreschkow for helpful discussion.
Parts of this research were supported by the Australian Research Council Centre of Excellence for All Sky Astrophysics in 3 Dimensions (ASTRO 3D), through project number CE170100013 .
Support for Program number HST-HF2-51406.001-A was provided by NASA through a grant from the Space Telescope Science Institute, which is operated by the Association of Universities for Research in Astronomy, Incorporated, under NASA contract NAS5-26555.


\appendix

\section{Equations for calculating neutral and molecular fractions}
\label{app:equations}
For completeness, here we compile the equations implemented in our code for calculating the neutral and molecular fractions of gas cells.  In principle, these can be used for any hydrodynamic simulation.  For the sake of brevity, we do not provide exhaustive descriptions of every variable in the equations that follow; for that, we refer the reader to the papers the prescriptions originate from.


\subsection{Neutral fractions of star-forming cells}
\label{app:neutral}
When a gas cell becomes sufficiently dense to form stars in TNG, it is modelled as a two-phase medium: one hot phase, one cold \citep{springel03}.  By including this sub-grid model, gas below the nominal simulation temperature floor (where metal line cooling is shut off) of $10^4$\,K can be probed.  We assume all gas in the cold phase of this model to be neutral.  Consistent with what is internally calculated in the TNG code, we therefore recover the neutral fraction of star-forming cells as
\begin{equation}
    f_n^{\rm SF} = \frac{u_{\rm hot} - u}{u_{\rm hot} - u_{\rm cold}}\,.
\end{equation}
Here, $u$ is the internal energy per unit mass of the cell, where $u_{\rm cold}$ and $u_{\rm hot}$ are equivalents for the hot and cold phases of each cell.  The cold phase is assumed to be an ideal gas at a temperature of 1000\,K.%
\footnote{In reality, neutral gas in galaxies can be cooler than this.  This assumption simply follows what was already implemented in TNG.  An investigation into the effect of the choices made in this sub-grid model on derived \HI~galaxy properties would be interesting, but that is beyond the scope of this paper.}
With the assumption that it is fully neutral, the mean molecular weight for this phase is
\begin{equation}
    \mu_{\rm cold} = \frac{4}{1 + 3\,X}\,,
\end{equation}
where $X$ is the hydrogen abundance fraction.%
\footnote{For $\mu_{\rm cold}$, $\mu_{\rm hot}$, and $\mu_4$, we actually approximate $X\!=\!0.76$, as per the initial conditions of TNG.  A more precise value of $X$ is entirely negligible here.  But for other calculations, we use the output hydrogen abundance for each cell from the simulation.}  
$u_{\rm hot}$ is dependent on the assumed temperature of a supernova ($T_{\rm SN}\!=\!5.73\!\times\!10^7$\,K) as follows:
\begin{subequations}
    \begin{equation}
        u_{\rm hot} \equiv u_{\rm cold} + \frac{u_{\rm SN}}{1+A}\,,
    \end{equation}
    \begin{equation}
        A \equiv A_0 \left(\frac{n_{\rm H}}{n_{\rm H,thresh}}\right)^{-0.8}\,,
    \end{equation}
\end{subequations}
where $A_0\!=\!573$ and $n_{\rm H}$ is the gas cell density in units of proton masses per cm$^3$. $n_{\rm H,thresh}$ is the threshold density for star formation, calculated as
\begin{equation}
	n_{\rm H,thresh} \equiv \frac{x_{\rm thresh}  \left[ \beta\, u_{\rm SN} - (1-\beta)\, u_{\rm cold} \right]}{\left( 1-x_{\rm thresh} \right)^{2} \left[ t_{*,0}\, X^2\, \Lambda(T) \right]}\,.
\end{equation}
TNG takes $\beta \! = \! 0.22578$ and $t_{*,0} \! = \! 3.27665\,{\rm Gyr}$.  $\Lambda(T)$ is the tabulated cooling function from \citet*{katz96}, and
\begin{equation}
    x_{\rm thresh} \equiv 1 + u_{\rm SN}^{-1}\, (1+ A_0)\, (u_{\rm cold} - u_4)\,.
\end{equation}
$u_4$ is the specific internal energy at 10\,000\,K.  As per the hot gas, this assumes full ionization, hence
\begin{equation}
    \mu_{\rm hot} = \mu_4 = \frac{4}{3+5\,X}\,.
\end{equation}

For non-star-forming cells, we use the internal neutral fractions already computed by the TNG model.  If our method were to be applied to a simulation where this field were not available for gas elements, one alternative would be to use the prescription of \citet{rahmati13a} in post-processing (but our tests found this would have produced notably different results for TNG).


\subsection{Molecular fractions of gas cells}
\label{app:molecular}

\subsubsection{GK11 prescription}
By running a series of `fixed ISM' hydrodynamic AMR simulations with detailed photo-chemical modelling, \citetalias{gk11} explored the dependence of the molecular fraction of gas cells on the local interstellar radiation field and dust-to-gas ratio. The outcome of their work was to provide fitting functions for the molecular fraction on these properties, which can be applied to simulations without the need for complex, explicit photo-chemical modelling.  Specifically, we apply equations 8, 10 \& 14 of \citetalias{gk11}:
\begin{subequations}
\begin{equation}
f_{\rm H_2} = \left( 1 + \frac{\Sigma_c}{\Sigma_{\rm H\,{\LARGE{\textsc i}}+H_2}} \right)^{-2}\,,
\end{equation}
\begin{equation}
\Sigma_c \equiv \frac{20\, \Lambda^{4/7}}{D_{\rm MW} \sqrt{1 + U_{\rm MW}\, D_{\rm MW}^2}}\, {\rm M}_{\odot}\,\rm{pc}^{-2}\,,
\end{equation}
\begin{equation}
\Lambda \equiv \ln\left[ 1 + g\, D_{\rm MW}^{3/7}\, \left(\frac{U_{\rm MW}}{15}\right)^{4/7} \right]\,,
\end{equation}
\begin{equation}
g \equiv \frac{1 + \alpha\, s + s^2}{1 + s}\,,
\end{equation}
\begin{equation}
s \equiv \frac{0.04}{D_* + D_{\rm MW}}\,,
\end{equation}
\begin{equation}
\alpha \equiv \frac{2.5\,U_{\rm MW}}{1 + (0.5\, U_{\rm MW})^2}\,,
\end{equation}
\begin{equation}
D_* \equiv 0.0015 \ln\left[ 1 + (3\, U_{\rm MW})^{1.7} \right]\,.
\end{equation}
\end{subequations}

\subsubsection{K13 prescription}
The prescription of \citetalias{k13} combines the work of \citet*{oml10} with the series of papers by \citet*{kmt08,kmt09,mckee10}.  \citet{oml10} described the cold ISM in two phases: one that clumps in clouds, and one that is diffuse.  From their model, they were able to derive the mid-plane pressure dependence of the molecular fraction presented by \citet{blitz04,blitz06}.   While \citet{oml10} explicitly state some of the cloud phase will be atomic, \citetalias{k13} approximated the clouds to be entirely molecular (with the diffuse cold gas as entirely atomic). 

We therefore apply equation 10 (and accompanying expressions) of \citetalias{k13} to TNG:
\begin{subequations}
\label{eq:k13}
\begin{equation}
f_{\rm H_2} = \left\{
\begin{array}{lr}
1 - 3\, S\, (4 + S)^{-1} & \forall  S<2\\
0 & \forall S \geq 2
\end{array}
\right.
\,,
\end{equation}
\begin{equation}
S \equiv \frac{\ln( 1 + 0.6\, \chi + 0.01\, \chi^2)}{0.6\, \tau_c}\,,
\end{equation}
\begin{equation}
\tau_c \equiv 0.066\, f_c\, D_{\rm MW}\, \frac{\Sigma_{\rm H\,{\LARGE{\textsc i}}+H_2}}{{\rm M}_{\odot}\,{\rm pc}^{-2}}\,,
\end{equation}
\begin{equation}
\chi \equiv 72\,U_{\rm MW} \left(\frac{n_{\rm CNM}}{{\rm cm}^{-3}}\right)^{-1}\,,
\end{equation}
\begin{equation}
n_{\rm CNM} \equiv {\rm max}[n_{\rm CNM,2p},\, n_{\rm CNM,hydro}]\,,
\end{equation}
\begin{equation}
n_{\rm CNM,2p} \equiv 23\, U_{\rm MW}\, \frac{4.1}{1 + 3.1\, D_{\rm MW}^{0.365}}\, {\rm cm}^{-3}\,,
\end{equation}
\begin{equation}
n_{\rm CNM,hydro} \equiv \frac{P_{\rm th}}{1.1\, k_B\, T_{\rm CNM,max}}\,,
\end{equation}
\begin{equation}
P_{\rm th} = \frac{\pi\, G\, \Sigma_{\rm H\,{\LARGE{\textsc i}}}^2}{4\, A} \left( 1 + R_{\rm H_2} + \sqrt{(1 + 2\, R_{\rm H_2})^2+\mathcal{F}} \right)\,,
\end{equation}
\begin{equation}
\mathcal{F} \equiv \frac{32\, \zeta_d\, A\, f_w\, c_w^2\, \rho_{\rm sd}}{\pi\, G\, \Sigma_{\rm H\,{\LARGE{\textsc i}}}^2}\,,
\end{equation}
\begin{equation}
R_{\rm H_2} \equiv \frac{f_{\rm H_2}}{1 - f_{\rm H_2}}\,,
\end{equation}
\end{subequations}
where $f_c \! = \! 5$ is the clumping factor, $T_{\rm CNM,max} \! = \! 243\,{\rm K}$ is the maximum temperature of the cold neutral medium, $A \! = \! 5$ is the relative pressure of turbulence and magnetic fields to thermal pressure, $c_w \! = \! 8\, {\rm km\, s}^{-1}$ is the sound speed of the warm neutral medium, $f_w \! = \! 0.5$, and $\zeta_d \! = \! 0.33$.  $\rho_{\rm sd}$ is the local density of stars and dark matter, which we calculate by subtracting the gas cell density from the total matter density found with a standard smoothed-particle hydrodynamics (SPH) kernel, accounting for all matter within a radius enclosing the 64 nearest dark-matter particles.  We note that \citet{oml10} actually define $\rho_{\rm sd}$ as the \emph{mid-plane} density of a galaxy \emph{disc}, and that their model for $n_{\rm CNM,hydro}$ was designed to describe the ISM.  In an approach where, a priori, we are agnostic about which gas medium (interstellar, circumgalactic, intrahalo, intergalactic, et cetera) a given cell belongs to, we have extrapolated the model to work on \emph{local} scales (i.e.~individual cells) for all gas media.  While perhaps not ideal, based on tests where we tried to separate the ISM and treated $\rho_{\rm sd}$ as a global galaxy property, the uncertainty this introduces to the \HI~masses of galaxies is small compared to the other assumptions made in our method.

Because $f_{\rm H_2}$ is solved for iteratively, the co-dependence of some of the expressions in Equation (\ref{eq:k13}) is not a problem.  Note that in their implementation of this prescription for EAGLE, the employed $P_{\rm th}$ expression by \citet{lagos15b} lacked the $R_{\rm H_2}$ terms and used $\Sigma_{\rm H\,{\LARGE{\textsc i}}+H_2}$ in place of $\Sigma_{\rm H\,{\LARGE{\textsc i}}}$ (therefore they did not iterate).  This has a non-negligible impact on the resulting $f_{\rm H_2}$, which may help to explain why we find less difference between prescriptions in this work.

\subsubsection{GD14 prescription}

The \citetalias{gd14} prescription updates that of \citetalias{gk11} by improving the modelling of self-shielding of H$_2$ and the employed cooling/heating functions.  \citetalias{gd14} offer two fitting functions for calculating molecular fractions, although their equation 6 is not appropriate for TNG, as it originates from very high-resolution simulations ($< \! 100\, {\rm pc}$ resolution).  Equation 8 of \citetalias{gd14} instead smooths over scales more comparable to TNG100's resolution.  This also applied to the \citetalias{gk11} prescription presented above.  The \citetalias{gd14} method we have applied to TNG100 is
\begin{subequations}
\begin{equation}
R_{\rm H_2} = \left( \frac{\Sigma_{\rm H\,{\LARGE{\textsc i}}+H_2}}{\Sigma_{R=1}} \right)^{\bar{\alpha}}\,,
\end{equation}
\begin{equation}
\Sigma_{R=1} = \frac{50\, \sqrt{0.001 + 0.1\,U_{\rm MW}}}{\bar{g}\, (1 + 1.69\, \sqrt{0.001 + 0.1\, U_{\rm MW}})}\, {\rm M}_{\odot}\, {\rm pc}^{-2}\,,
\end{equation}
\begin{equation}
\bar{\alpha} \equiv 0.5 + \left(1 + \sqrt{\frac{U_{\rm MW}\, D_{\rm MW}^2}{600}} \right)^{-1}\,,
\end{equation}
\begin{equation}
\bar{g} \equiv \sqrt{\bar{D}_*^2 + D_{\rm MW}^2}\,,
\end{equation}
\begin{equation}
\bar{D}_* \equiv 0.17\, \frac{2 + \bar{S}^5}{1 + \bar{S}^5}
\end{equation}
\end{subequations}
\citepalias[note the update to this in the erratum of][]{gd14}.  We have adopted the effective cell length for $\bar{S}$ [i.e.~$\bar{S} \! = \! (m / \rho)^{1/3} / (100\,{\rm pc})$ for a gas cell].  As per \citet{lagos15b}, if this prescription were to instead be applied to an SPH simulation (where there is no direct equivalent for a cell length), $\bar{S}$ could be approximated as the Jeans length (Equation \ref{eq:jeans}).

\section{Resolution test}
\label{app:res}

\begin{figure}
\centering
\includegraphics[width=\textwidth]{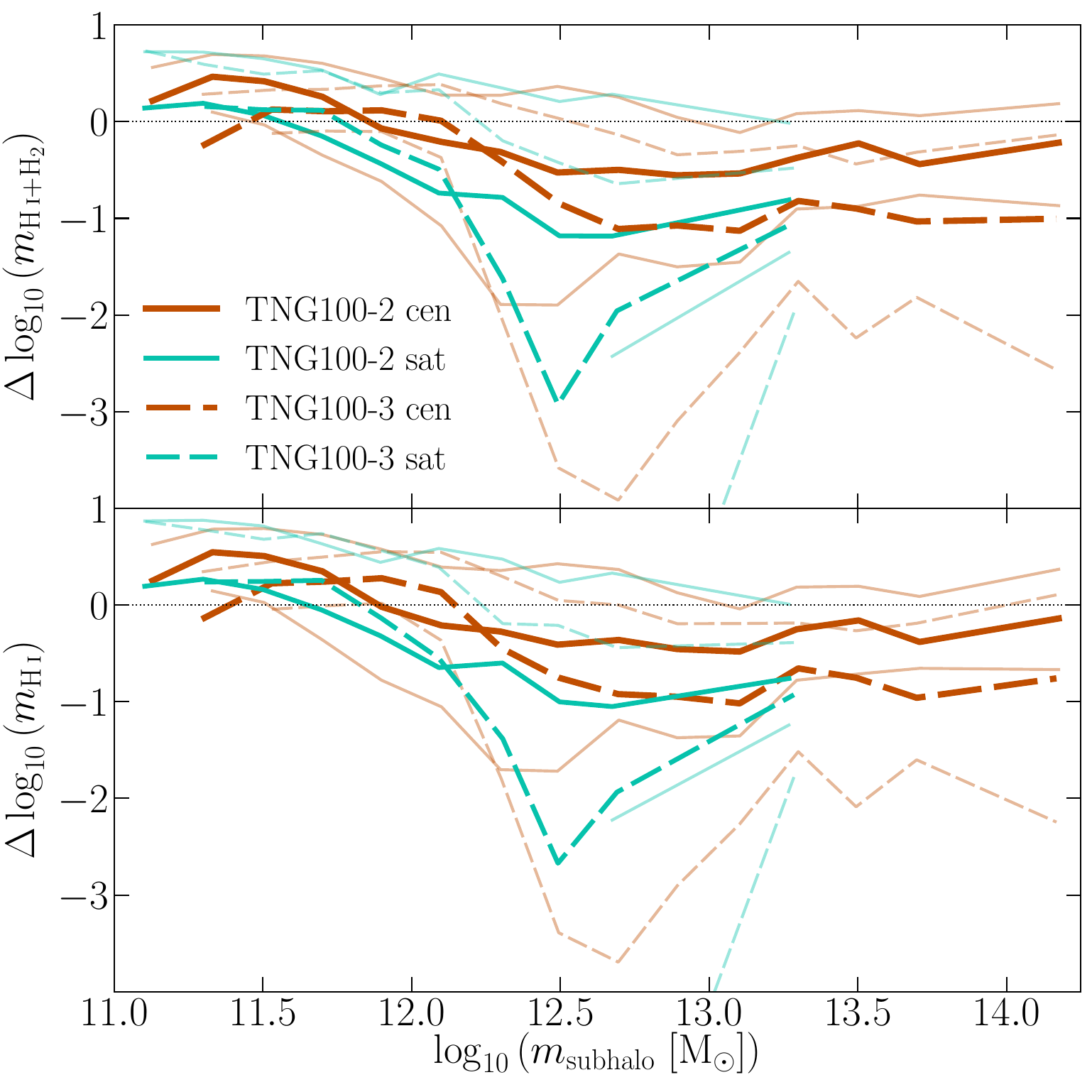}
\caption{Difference in neutral-hydrogen mass (top panel) and \HI~mass (bottom panel, assuming the \citetalias{gd14} prescription) in TNG runs of lower resolution compared to the main TNG100 run, as a function of subhalo mass.  Thick lines compare the medians from TNG100-2 and TNG100-3 to that of TNG100-1.  Thin lines compare the $16^{\rm th}$ and $84^{\rm th}$ percentiles of TNG100-2 and TNG100-3 to the median of TNG100-1.}
\label{fig:res}
\end{figure}

We would be remiss if we did not mention how numerically converged our results are with the mass resolution of the simulation.  While a full analysis of numerical convergence is well beyond the scope of this paper, to give some indication of its significance to our results, we show in Fig.~\ref{fig:res} the change in gas mass (first accounting exclusively for neutral hydrogen and then exclusively just \HI) as a function of subhalo mass for runs of TNG100 at lower resolution.  The particle mass in TNG100-2 is eight times greater than TNG100-1 (i.e.~what has just been referred to as TNG100 throughout this paper), with particles in TNG100-3 a further eight times more massive than that \citep[see table 1 of][]{pillepich18b}.  

Using subhalo mass on the $x$-axis rather than, e.g., stellar mass reduces the effect that resolution has on this axis, making the $y$-axis the primary direction for resolution to cause variations.  In truth though, (sub)halo mass has a dependence on feedback strength \citep[e.g.][]{schaller15a}, which itself is resolution-dependent, so there is still an effect on both axes.

Unsurprisingly, the neutral-gas content of galaxies in TNG is not converged with mass resolution; this is generally true for galaxy properties in hydrodynamic simulations.  Where Fig.~\ref{fig:res} offers some solace is that the variation in \HI~mass tracks the variation in neutral mass almost exactly.  This means any resolution dependence of our post-processing method for the \HI/\Htwo~breakdown (Section \ref{ssec:hih2}) is entirely negligible for our \HI~results (this does not mean the same is true for \Htwo~properties of galaxies).

Interestingly, the subhalo mass scale where there is the greatest resolution dependence ($\sim\!10^{12.5}\,{\rm M}_{\odot}$) corresponds to the same mass scale where the dip in neutral fraction in the left panel of Fig.~\ref{fig:NeutralFrac} is seen.

We note that TNG100-2 and TNG100-3 were \emph{not} recalibrated.  Comparing recalibrated simulations would be another valuable form of resolution test \citep[e.g.~as discussed by][]{schaye15}.  But because neutral-gas content did not feature in TNG's calibrations, there is no guarantee this would lead to convergence in this property.  Indeed, as has been shown for the EAGLE simulations, higher-resolution, recalibrated runs can show notably different \HI~statistics for galaxies \citep[see][]{bahe16,marasco16,crain17}.

More on the numerical convergence of our methods can be found in appendix B of \citet{diemer18}.

\label{lastpage}
\end{document}